\documentclass[twocolumn,showpacs,preprintnumbers,amsmath,amssymb]{revtex4}

\usepackage{graphicx}
\usepackage{dcolumn}
\usepackage{bm}

\begin{document}

\preprint{Phys. Rev. E (in print)}

\title{Effect of mixing and spatial dimension on the glass transition}

\author{David Hajnal$^1$}\email{hajnalda@uni-mainz.de}
\author{Joseph M. Brader$^2$}
\author{Rolf Schilling$^1$}
\affiliation{$^1$Institut f\"ur Physik, Johannes Gutenberg-Universit\"at Mainz, Staudinger Weg 7, D-55099 Mainz, Germany\\
$^2$Fachbereich Physik, Universit\"at Konstanz, D-78457 Konstanz, Germany}

\date{\today}

\begin{abstract}
We study the influence of composition changes on the glass transition
of binary hard disc and hard sphere mixtures in the framework of mode
coupling theory. We derive a general expression for the slope of a glass
transition line. Applied to the binary mixture in the low
concentration limits, this new method allows a fast prediction
of some properties of the glass transition lines.
The glass transition diagram we find for binary hard discs
strongly resembles the random close packing diagram.
Compared to 3D from previous studies, the extension of the glass regime due to mixing is much more
pronounced in 2D where plasticization only sets in at larger size disparities.
For small size disparities we find a stabilization of the glass phase quadratic in the deviation
of the size disparity from unity.
\end{abstract}


\maketitle

\section{Introduction}

Adding a second component to a one-component liquid changes its
static and dynamical properties.
For instance,
if one adds a low concentration of rather small species, depletion forces between the larger particles are induced
\cite{asakura54}.
These effective forces are attractive for small separations and tend
to stabilize the liquid phase in addition to influencing the transport
properties \cite{Lue02,Horbach02,Voigtmann06}.
Such mixing effects are interesting from a fundamental point of view, but also for applications. It is our main goal
to study the influence of mixing on the glass transition of binary systems with hard core interactions in two and three
dimensions. This will be done in the framework of mode coupling theory (MCT) \cite{Goetze09}.

Mixing effects on the MCT glass transition were studied first by Barrat and Latz \cite{Barrat90} for binary soft spheres.
However, the first \textit{systematic} investigation was performed by G\"otze and Voigtmann \cite{Voigtmann03}
for binary hard spheres with moderate size ratios $\delta=R_s/R_b$. $R_s$ and $R_b$ are the radii of the small
and big spheres, respectively. For size ratios close to unity, a slight extension of the glass regime was observed.
Larger size disparities induce a plasticization effect, leading to a stabilization of the liquid due to mixing.
The results qualitatively agree with those from dynamic light scattering experiments \cite{Henderson96,Williams01}
and molecular dynamics simulations \cite{Foffi03,Foffi04}.
In contrast to this, a recent theory of Ju\'arez-Maldonado and Medina-Noyola based on the
self consistent generalized Langevin equation (SCGLE) \cite{Juarez08} predicts a plasticization effect also
for size ratios close to unity.
These authors argue that the data available from simulations and experiments
are not sufficiently accurate to rule out one of the scenarios.
Size ratios far from unity, i.e. $\delta\ll 1$, may be problematic. First, the quality of e.g. Percus-Yevick (PY) theory
used to calculate the static input for MCT
may become
less reliable.
Second, phase separation (see discussion in Ref.~\cite{goetzelmann98})
and third, a sequential arrest of the big and small particles (by a type-A transition) could occur.
The diverging lengthscale associated with a type-A transition affects the quality of the MCT approximations.

The results of G\"otze and Voigtmann \cite{Voigtmann03} exhibit four mixing effects,
two of which were mentioned above.
The two remaining mixing effects are an increase of the plateau values of the normalized correlation functions for intermediate times
for almost all wave numbers upon increasing the concentration of the smaller particles,
and a slowing down of the initial part of the relaxation of the big-big correlators towards these plateaus.
Our motivation is twofold. \textit{First}, we
want to explore whether these effects also exist in a corresponding \textit{two-dimensional} liquid of binary hard discs.
A recent experiment \cite{Koenig05} has given evidence for glassy behavior in a similar two-dimensional liquid
including dipolar interactions. \textit{Second}, we will investigate in more detail the influence of mixing
\textit{close} to the monodisperse system, i.e.
fixing the packing fraction $\varphi$ at $\varphi_0^c$ (the critical packing fraction of the \textit{monodisperse} system), how does
a very small perturbation of the \textit{monodisperse} system influence the glass transition? The arbitrary small perturbation
can be achieved in three ways,
either by adding a very small concentration of smaller or bigger species for given arbitrary $\delta<1$, or by a slight decrease of the
diameter of an arbitrary concentration $x_s$ of the smaller particles, accompanied by a slight increase of the remaining particles,
i.e. $1-\delta\ll1$.

The mixing effects in the low concentration limits follow directly from the slopes $\partial\varphi^c(x_s,\delta)/\partial x_s$ at $x_s=0$ and $x_s=1$ of the glass transition lines (GTLs) $\varphi^c(x_s,\delta)$ at fixed $\delta$.
If $\partial\varphi^c(x_s,\delta)/\partial x_s|_{x_s=0}$ is positive (negative), the liquid (glass) is stabilized.
The same is true if $\partial\varphi^c(x_s,\delta)/\partial x_s|_{x_s=1}=-\partial\varphi^c(1-x_b,\delta)/\partial x_b|_{x_b=0}$
is negative (positive). $x_b=1-x_s$ is the big particle's concentration. 
Since the determination of these slopes from the numerical result for $\varphi^c(x_s,\delta)$ with
discretized values of $x_s$ is not precise, particularly for $\delta$ closer to unity (cf. the critical lines for $\delta=0.7$
and $0.8$ in Fig.~1 of Ref.~\cite{Voigtmann03}), we will derive an analytical expression for
$\partial\varphi^c(x_s,\delta)/\partial x_s$
for arbitrary $x_s$ and $\delta$. Applied to $x_s=0$ and $x_s=1$, only the glass transition singularity of the monodisperse system
is needed. The remaining quantities entering the slope at $x_s=0$ and $x_s=1$ can be determined from a perturbational approach discussed below. The application for the slope formula will be done for both, hard discs and hard spheres. This allows to explore the
dimensional dependence (at least for $d=2$ and $d=3$) of the mixing effects in the weak mixing limit.

\section{Mode coupling theory}

We will restrict ourselves to the essential equations to keep our presentation self-contained. For details, the reader may consult
Ref.~\cite{Goetze09}.
Correlation functions are matrix valued vectors denoted by bold symbols $\boldsymbol{A}$, $\boldsymbol{B}$ etc.
Their components $\boldsymbol{A}_k$, $\boldsymbol{B}_k$ being $M\times M$ matrices $(A_k^{\alpha\beta})$, $(B_k^{\alpha\beta})$
(in case of an $M$-component fluid) are labelled by subscript Latin indices (the wave numbers)
which can be taken from a discrete or a continuous set.
The elements $A_k^{\alpha\beta}$, $B_k^{\alpha\beta}$ of these matrices are indicated by superscript Greek indices,
in some cases these elements shall also be denoted by $(\boldsymbol{A})_k^{\alpha\beta}$, $(\boldsymbol{B})_k^{\alpha\beta}$.
Matrix products are defined componentwise, i.e. $\boldsymbol{C}=\boldsymbol{A}\boldsymbol{B}$ reads $\boldsymbol{C}_k=\boldsymbol{A}_k\boldsymbol{B}_k$ for all $k$.
We call $\boldsymbol{A}$ positive (semi-)definite,
($\boldsymbol{A}\succeq\boldsymbol{0}$), $\boldsymbol{A}\succ\boldsymbol{0}$ if this is true for all $\boldsymbol{A}_k$.
$\boldsymbol{0}$ denotes the (generalized) zero matrix.
If $k$ is restricted to a finite number of values, then the
standard scalar product of $\boldsymbol{A}$ and $\boldsymbol{B}$ shall be defined as $(\boldsymbol{A}|\boldsymbol{B})=\sum_k\sum_{\alpha,\beta}A_k^{\alpha\beta}B_k^{\alpha\beta}$.

\subsection{General equations}

We consider an isotropic and homogeneous classical fluid consisting of $M$ macroscopic components in $d$ dimensions. $\boldsymbol{\Phi}(t)$ denotes the matrix of time dependent partial autocorrelation functions of density fluctuations,
$\Phi_k^{\alpha\beta}(t)$ ($\alpha,\beta=1,\dots,M$), at wave number $k$. We require the normalization $\boldsymbol{\Phi}(0)=\boldsymbol{S}$, where $\boldsymbol{S}$ denotes the static structure factor matrix whose elements obey
$\lim_{k\rightarrow\infty}S_{k}^{\alpha\beta}=x_{\alpha}\delta_{\alpha\beta}$.
Here $\delta_{\alpha\beta}$ denotes
the Kronecker delta and $x_{\alpha}$ the particle number concentration of
component $\alpha$.

Considering overdamped colloidal dynamics, the Zwanzig-Mori projection operator formalism yields the equation of motion
\begin{equation}
\label{zwanzig_mori}
\boldsymbol{\tau}\boldsymbol{\dot{\Phi}}(t)+\boldsymbol{S}^{-1}\boldsymbol{\Phi}(t)+\int_0^t\mathrm{d} t'\boldsymbol{m}(t-t')\boldsymbol{\dot{\Phi}}(t')=\boldsymbol{0}
\end{equation}
with the memory kernel $\boldsymbol{m}(t)$ describing fluctuating stresses and playing the role of generalized friction.
$\boldsymbol{\tau}$ is a positive definite matrix of microscopic relaxation times.
Its components shall be approximated by
$\tau_k^{\alpha\beta}=\delta_{\alpha\beta}/(k^2D^0_{\alpha}x_{\alpha})$
where hydrodynamic interactions are neglected. $D^0_{\alpha}$ denotes the short-time diffusion coefficient of a single
particle of the species $\alpha$ inserted into the fluid.
With this, the short-time asymptote of $\boldsymbol{\Phi}(t)$ is given by
\begin{equation}
\label{short-time}
\boldsymbol{\Phi}(t\rightarrow 0)=\boldsymbol{S}-\boldsymbol{\tau}^{-1}t+\mathcal{O}(t^2),
\end{equation}
here we restrict ourselves to $t\geq0$.
MCT approximates $\boldsymbol{m}(t)$ by a symmetric bilinear functional
$\boldsymbol{\mathcal{F}}$ of $\boldsymbol{\Phi}(t)$,
\begin{equation}
\boldsymbol{m}(t)=\boldsymbol{\mathcal{F}}[\boldsymbol{\Phi}(t),\boldsymbol{\Phi}(t)].
\end{equation}
It is straightforward to generalize the explicit expression for $\boldsymbol{\mathcal{F}}$ of a simple fluid in $d$ dimensions
($d\geq2$) presented in Ref.~\cite{Bayer07} to multicomponent systems. The result is
\begin{eqnarray}
\label{memory}
\mathcal{F}_k^{\alpha\beta}[\boldsymbol{X},\boldsymbol{Y}]&=&\frac{\Omega_{d-1}}{(4\pi)^d}
\sum_{\alpha',\beta',\alpha'',\beta''}\int_0^\infty \mathrm{d} p\int_{|k-p|}^{k+p} \mathrm{d} q\nonumber\\
&&\times V^{\alpha\beta;\alpha'\beta',\alpha''\beta''}_{k;p,q}X_p^{\alpha'\beta'}Y_q^{\alpha''\beta''}
\label{bilinear}
\end{eqnarray}
with the vertices
\begin{equation}
\label{vertex_1}
V^{\alpha\beta;\alpha'\beta',\alpha''\beta''}_{k;p,q}=
\frac{n}{x_{\alpha}x_{\beta}}\frac{pq}{k^{d+2}}v_{kpq}^{\alpha\alpha'\alpha''}v_{kpq}^{\beta\beta'\beta''}
\end{equation}
where
\begin{equation}
\label{vertex}
v^{\alpha\beta\gamma}_{kpq}=\frac{(k^2+p^2-q^2)c_p^{\alpha\beta}\delta_{\alpha\gamma}
+(k^2-p^2+q^2)c_q^{\alpha\gamma}\delta_{\alpha\beta}}{[
4k^2p^2-(k^2+p^2-q^2)^2]^{(3-d)/4}}.
\end{equation}
$c_k^{\alpha\beta}$ denote the direct correlation functions.
$\boldsymbol{c}$ is related to $\boldsymbol{S}$ via the Ornstein-Zernike (OZ) equation
\begin{equation}
\label{oz1}
(\boldsymbol{S}^{-1})_k^{\alpha\beta}=\delta_{\alpha\beta}/x_{\alpha}-nc_k^{\alpha\beta}.
\end{equation}
$n$ is the total number of particles per volume and
$\Omega_d=2{\pi}^{d/2}/\Gamma(d/2)$ the well known result for the surface of a unit sphere in $d$ dimensions. $\Gamma(x)$ is
the gamma function.

\subsection{Definition of the model}

The $M$-component MCT in $d$ dimensions shall be applied to binary hard ``sphere'' mixtures (HSM) in $d$ dimensions consisting
of big ($\alpha=b$) and small ($\alpha=s$) particles.
Let $R_{\alpha}$ denote the radius of the species $\alpha$.
Three independent control parameters are necessary to characterize the thermodynamic state of a HSM.
We choose them to be the total packing fraction $\varphi=\varphi_s+\varphi_b$ with
$\varphi_{\alpha}=nx_{\alpha}(\Omega_d/d)R_{\alpha}^d$, the size ratio $\delta=R_s/R_b\leq1$,
and the particle number concentration $x_s$ of the smaller particles.

For the following, we discretize the MCT equations, i.e. $k$ is discretized to a
finite, equally spaced grid of $K$ points, $k=(\hat{o}_d+\hat{k})\Delta k$ with $\hat{k}=0,1,\dots,K-1$ and $0<\hat{o}_d<1$.
The integrals in Eq.~(\ref{memory}) are then replaced by Riemann sums
\begin{equation}
\label{riemann}
\int_0^\infty \mathrm{d} p\int_{|k-p|}^{k+p} \mathrm{d} q \dots \mapsto (\Delta k)^2 \sum_{\hat{p}=0}^{K-1}  \sum_{\hat{q}=|\hat{k}-\hat{p}|}^{\min\{K-1,\hat{k}+\hat{p}\}}\dots
\end{equation}
and Eq.~(\ref{zwanzig_mori}) represents a
finite number of coupled nonlinear ``integro''-differential equations.
We further restrict our numerical studies to the cases $d=2$ and $d=3$. For the offset, following previous works, 
we choose $\hat{o}_2=0.303$ for $d=2$ \cite{Bayer07} 
and $\hat{o}_3=0.5$ for $d=3$ \cite{Voigtmann03}. The choice $K=250$ and $\Delta k=0.3$
turns out to be sufficiently accurate to avoid larger discretization effects.

For calculations with finite concentrations of both particle species, the unit length
shall be given by the diameter $2R_b$ of the bigger particles,
and the short-time diffusion coefficients $D^0_{\alpha}$ shall be assumed to obey the Stokes-Einstein law.
Further, the unit of time is chosen such that $D^0_{\alpha}=0.01/(2R_{\alpha})$.
For the numerical solution of Eq.~(\ref{zwanzig_mori}) we use the algorithm first published in
\cite{Hofacker91}. Our time grids consist of $256$ points, as initial step size we choose $10^{-8}$ time units.

For the discussion of the weak mixing limits (see below) it is convenient to choose the
diameters $2R_{\alpha}$ of the majority particle species as unit length.

\subsection{Static structure}

Approximate closures of the OZ equation
provide the most powerful methods currently available for a fast
calculation of the pair correlation functions from first principles \cite{hansen}.
The OZ equation for an arbitrary mixture is given by
\begin{equation}
\boldsymbol{h} = \boldsymbol{c}+n\boldsymbol{c}\boldsymbol{x}\boldsymbol{h}
\label{oz}
\end{equation}
where $x_k^{\alpha\beta}=x_{\alpha}\delta_{\alpha\beta}$ and the $h^{\alpha\beta}_k$ are the total correlation functions.
For our binary HSM model we use the PY approximation given by
\begin{equation}
\begin{array}{rcrr}
h^{\alpha\beta}(r)&=&-1,& r<(R_{\alpha}+R_{\beta}),\\
c^{\alpha\beta}(r)&=&0,& r>(R_{\alpha}+R_{\beta}).
\end{array}
\label{py}
\end{equation}
In odd dimensions the coupled Eqs.~(\ref{oz}) and (\ref{py}) can be
solved analytically \cite{Lebowitz}. In even dimensions numerical methods must be employed.
Among the several existing algorithms \cite{brader_oz}
we use the classical Lado algorithm \cite{lado} for simplicity.
In our numerical solution of the 2D system we use a real space cutoff $r_{max}=50$ with $4000$ grid points.

\subsection{Glass transition lines}

The nonergodicity parameters (NEPs) $\boldsymbol{F}=(F_k^{\alpha\beta})$ are given by
$\boldsymbol{F}=\lim_{t\rightarrow\infty}\boldsymbol{\Phi}(t)$.
For the discretized model described above, the following statements can
be proved \cite{Franosch02}. Equation~(\ref{zwanzig_mori}) has a unique solution. It
is defined for all $t\geq0$ and is completely monotone, $(-\partial/\partial_t)^n\boldsymbol{\Phi}(t)\succeq\boldsymbol{0}$.
$\boldsymbol{F}\succeq\boldsymbol{0}$ is (with respect to $\succeq$) the maximum real, symmetric fixed point of
the nonlinear map
\begin{equation}
\label{fixed_point}
\boldsymbol{\mathcal{I}}[\boldsymbol{X}]=\boldsymbol{S}-(\boldsymbol{S}^{-1}+\boldsymbol{\mathcal{F}}[\boldsymbol{X},\boldsymbol{X}])^{-1}.
\end{equation}
Iterating Eq.~(\ref{fixed_point}) starting with $\boldsymbol{X}=\boldsymbol{S}$ leads to a monotonically decaying sequence converging
towards $\boldsymbol{F}$.
Linearization of $\boldsymbol{\mathcal{I}}$ yields
the positive definite linear map (stability matrix)
\begin{equation}
\label{stability}
\boldsymbol{C}[\boldsymbol{Y}]=2(\boldsymbol{S}-\boldsymbol{F})\boldsymbol{\mathcal{F}}[\boldsymbol{F},\boldsymbol{Y}](\boldsymbol{S}-\boldsymbol{F})
\end{equation}
with $\boldsymbol{C}[\boldsymbol{Y}]\succeq\boldsymbol{0}$ for all $\boldsymbol{Y}\succeq\boldsymbol{0}$.
From a physical point of view, it is reasonable to assume that $\boldsymbol{C}$ is irreducible if $\boldsymbol{F}\succ\boldsymbol{0}$
\cite{Franosch02}.
$\boldsymbol{C}$
has then a nondegenerate maximum
eigenvalue $0<r\leq1$
with a corresponding (right) eigenvector $\boldsymbol{H}\succ\boldsymbol{0}$.
For any other eigenvalue $\tilde{r}$ of $\boldsymbol{C}$, $|\tilde{r}|\leq|r|$ holds, and if $|\tilde{r}|=|r|$, then
the corresponding eigenvector can not be positive definite.
Hence, possible MCT singularities are identified by $r=1$ and belong to the class $A_l,$ $l=2,3,\dots,$ introduced by
Arnol'd \cite{Arnold}.
The adjoint map $\hat{\boldsymbol{C}}$ of $\boldsymbol{C}$ satisfies
$(\hat{\boldsymbol{C}}[\boldsymbol{A}]|\boldsymbol{B})=
(\boldsymbol{A}|\boldsymbol{C}[\boldsymbol{B}])
$ for all $\boldsymbol{A}$, $\boldsymbol{B}$. Its eigenvector $\hat{\boldsymbol{H}}\succ\boldsymbol{0}$ is the left eigenvector of $\boldsymbol{C}$ corresponding to the eigenvalue $r$.
These two eigenvectors are determined uniquely by requiring the normalization
\begin{equation}
\label{norm}
(\hat{\boldsymbol{H}}|\boldsymbol{H})=(\hat{\boldsymbol{H}}|\boldsymbol{H}\{\boldsymbol{S}-\boldsymbol{F}\}^{-1}\boldsymbol{H})=1.
\end{equation}

For binary HSM models, higher order singularities may occur for large size disparities where the packing
contributions $\hat{x}_{\alpha}=\varphi_{\alpha}/\varphi$ of both components are of the same order \cite{Voigtmann08}.
In the present paper,
we restrict our discussion to the generic (type-B) MCT bifurcations belonging to the class $A_2$
where
$\boldsymbol{F}$ jumps from $\boldsymbol{0}$ to $\boldsymbol{F}^c\succ\boldsymbol{0}$.
Quantities taken at critical points shall be indicated by a superscript $c$.
The glass transition takes place at the critical surface $\varphi^c(x_s,\delta)$ within the three-dimensional physical
parameter space $(\varphi,x_s,\delta)$. $\varphi^c(x_s,\delta)$ fulfills \footnote{Here we allow that $\delta$ varies between
zero and infinity. If $\delta>1$, then $x_s$ plays the role of the concentration of the big particles.}
\begin{equation}
\label{symmetry}
\varphi^c(x_s,\delta)=\varphi^c(1-x_s,1/\delta).
\end{equation}
Equation~(\ref{symmetry}) demonstrates that $\varphi^c(x_s,\delta)$ for fixed $\delta$ is not symmetric with respect to
the equimolar concentration $x_s=1/2$. However, for $x_s=1/2$ and small disparity, i.e. $\varepsilon=(1-\delta)\ll1$,
it follows from Eq.~(\ref{symmetry}) that $\varphi^c(1/2,\delta)\sim\varepsilon^2$ in leading order in $\varepsilon$.
Accordingly, for the equimolar situation and small disparity the influence of disparity is quadratic only.
$\varphi^c(x_s,\delta)$ can be determined numerically by a simple bisection algorithm monitoring the NEPs.

\subsection{Slope of a critical line}

For a general model system with $L$ external, i.e. physical control parameters $\vec{\xi}=(\xi_1,\dots,\xi_L)$,
the generic glass transition singularities form a $(L-1)$-dimensional hypersurface $\mathcal{H}$.
Locally, this surface can be represented, e.g. as $\xi_l^c(\xi_1,\dots,\xi_{l-1},\xi_{l+1},\dots,\xi_L)$ for any $l$. For fixed
$\xi_i$, $i\neq j$, $i\neq l$,
$\xi_l^c(\xi_1,\dots,\xi_j,\dots,\xi_{l-1},\xi_{l+1},\dots,\xi_L)$ describes a GTL which is a function of $\xi_j$.
An expression for its slope
$(\partial\xi_l^c/\partial\xi_j)(\xi_1,\dots,\xi_j,\dots,\xi_{l-1},\xi_{l+1},\dots,\xi_L)$
is obtained
by use of the separation parameter $\sigma$. Let $\vec{\xi}^c\in\mathcal{H}$ be a critical point and
$\Delta\vec{\xi}=\vec{\xi}-\vec{\xi}^c$. Then the separation parameter is a linear function $\sigma(\Delta\vec{\xi})$ in $\Delta\vec{\xi}$ \cite{Goetze09}. $\sigma(\Delta\vec{\xi})=0$ defines the tangent plane of the hypersurface
$\mathcal{H}$ at the critical point $\vec{\xi}^c$.
Then it is easy to prove that
\begin{eqnarray}
\label{slope1}
(\partial\xi_l^c/\partial\xi_j)(\xi_1^c,\dots,\xi_j^c,\dots,\xi_{l-1}^c,\xi_{l+1}^c,\dots,\xi_L^c)\nonumber\\
=
-\left.\frac{\partial\sigma/\partial(\Delta\xi_j)}{\partial\sigma/\partial(\Delta\xi_l)}\right|_{\Delta\vec{\xi}=\vec{0}}.
\end{eqnarray}
The separation parameter $\sigma(\Delta\vec{\xi})$ follows from
\begin{eqnarray}
\label{sigma}
\tilde{\sigma}(\vec{\xi})&=&(\hat{\boldsymbol{H}}^c|\{\boldsymbol{S}^c-\boldsymbol{F}^c\}{\boldsymbol{S}^c}^{-1}
\{\boldsymbol{S}\boldsymbol{\mathcal{F}}[\boldsymbol{F}^c,\boldsymbol{F}^c](\boldsymbol{S}-\boldsymbol{F}^c)\nonumber\\
&&-\boldsymbol{S}^c\boldsymbol{\mathcal{F}}^c[\boldsymbol{F}^c,\boldsymbol{F}^c](\boldsymbol{S}^c-\boldsymbol{F}^c)\})
\end{eqnarray}
by expanding around $\vec{\xi}^c$ up to linear order in $\Delta\vec{\xi}$ \cite{Goetze09,Voigtmann_PhD}.
The result~(\ref{slope1}) demonstrates that the separation parameter besides being a measure for the distance from the critical
point $\vec{\xi}^c$ also contains local information of a GTL.

Applied to a binary liquid, Eq.~(\ref{slope1}) yields
\begin{equation}
\label{slope}
\left.\frac{\partial \varphi^c}{ \partial x_s}\right|_{(x_s,\delta)=(x_s^c,\delta^c)}=-\left.\frac{\partial\sigma/\partial (\Delta x_s)}{\partial\sigma/\partial (\Delta\varphi)}\right|
_{(\Delta\varphi,\Delta x_s,\Delta\delta)=\vec{0}}
\end{equation}
where $\Delta\delta$ and all critical input parameters have to be considered as fixed constants when calculating the
partial derivatives. A similar expression follows for $(\partial\varphi^c/\partial\delta)(x_s=x_s^c,\delta=\delta^c)$.

Let us further remark that the concept of introducing a separation parameter is not restricted to MCT models.
Thus, Eq.~(\ref{slope1}) holds for \textit{any} system which has at least
two control parameters and exhibits the generic $A_2$ bifurcation scenario \cite{Arnold}.

\subsection{Weak mixing limit}

One of the central aspects of our paper is to demonstrate the 
predictive power of Eq.~(\ref{slope}) for the limits $x_s=0$ and $x_b=0$.
By performing these limits analytically, we obtain formulae whose numerical evaluation is much less time consuming then
the numerical procedure mentioned above, i.e. to determine the slope from $\varphi^c(x_s,\delta)$. Note, the knowledge of the initial slopes of the GTLs for both limits is already
sufficient to estimate their qualitative behavior under certain assumptions.
The essential steps for the calculation of the slope $\partial\varphi^c/\partial x_s$ are explained in Appendix~\ref{weakmixing}.

\section{Results and discussion}

\subsection{Glass transition lines}

\begin{figure}
\includegraphics[width=1\columnwidth]{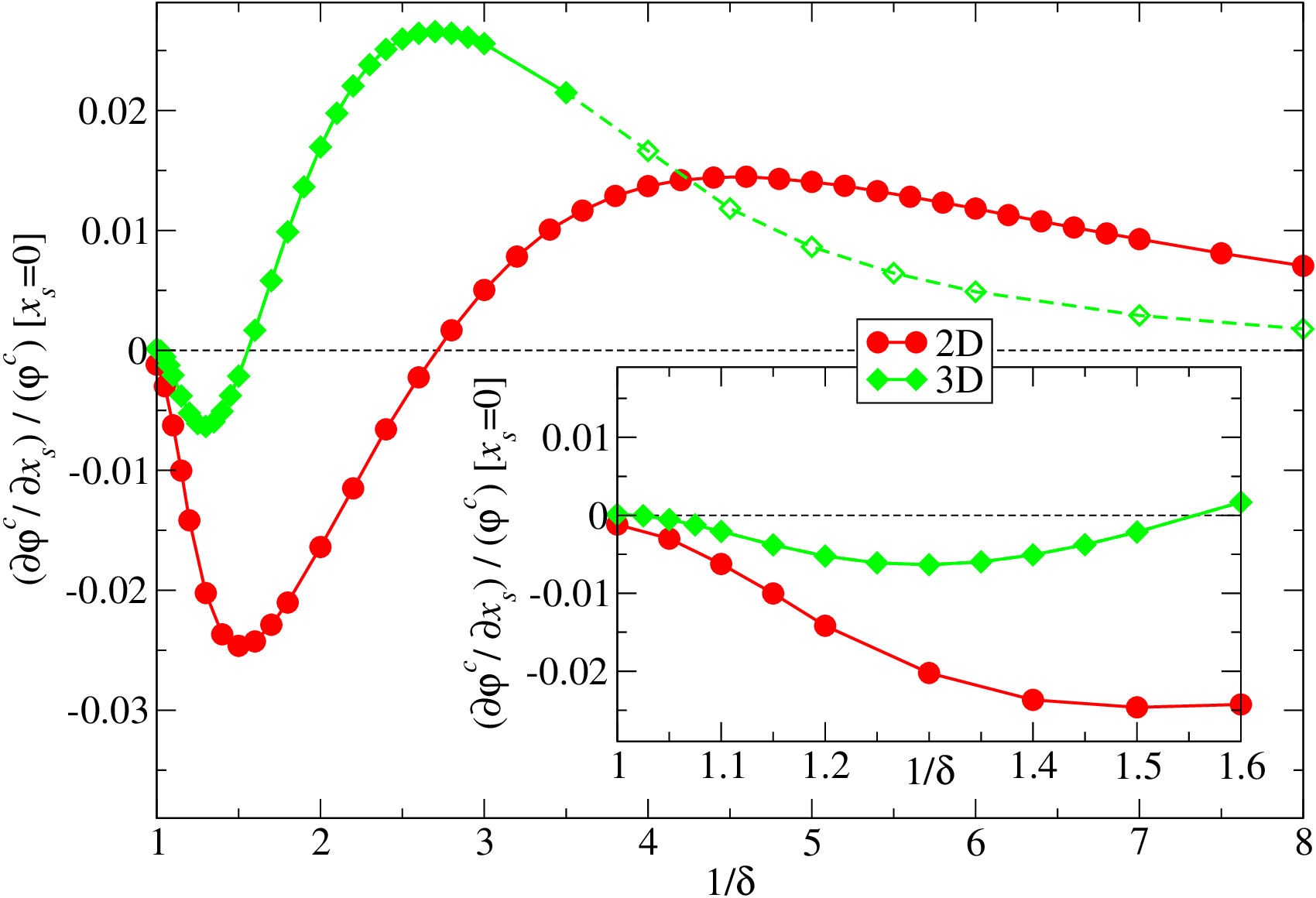}
\caption{(Color online) Normalized slopes of the GTLs at $x_s=0$ for the binary HSM models in 2D and 3D.
It is $\varphi^c_0\cong0.6914$ for $d=2$ and $0.5159$ for $d=3$.
For $1/\delta>3.5$ in the 3D model, the tagged particle NEPs indicate a
delocalization transition of the smaller spheres. This regime is indicated by open symbols (see text).}
\label{Fig.1}
\end{figure}

\begin{figure}
\includegraphics[width=1\columnwidth]{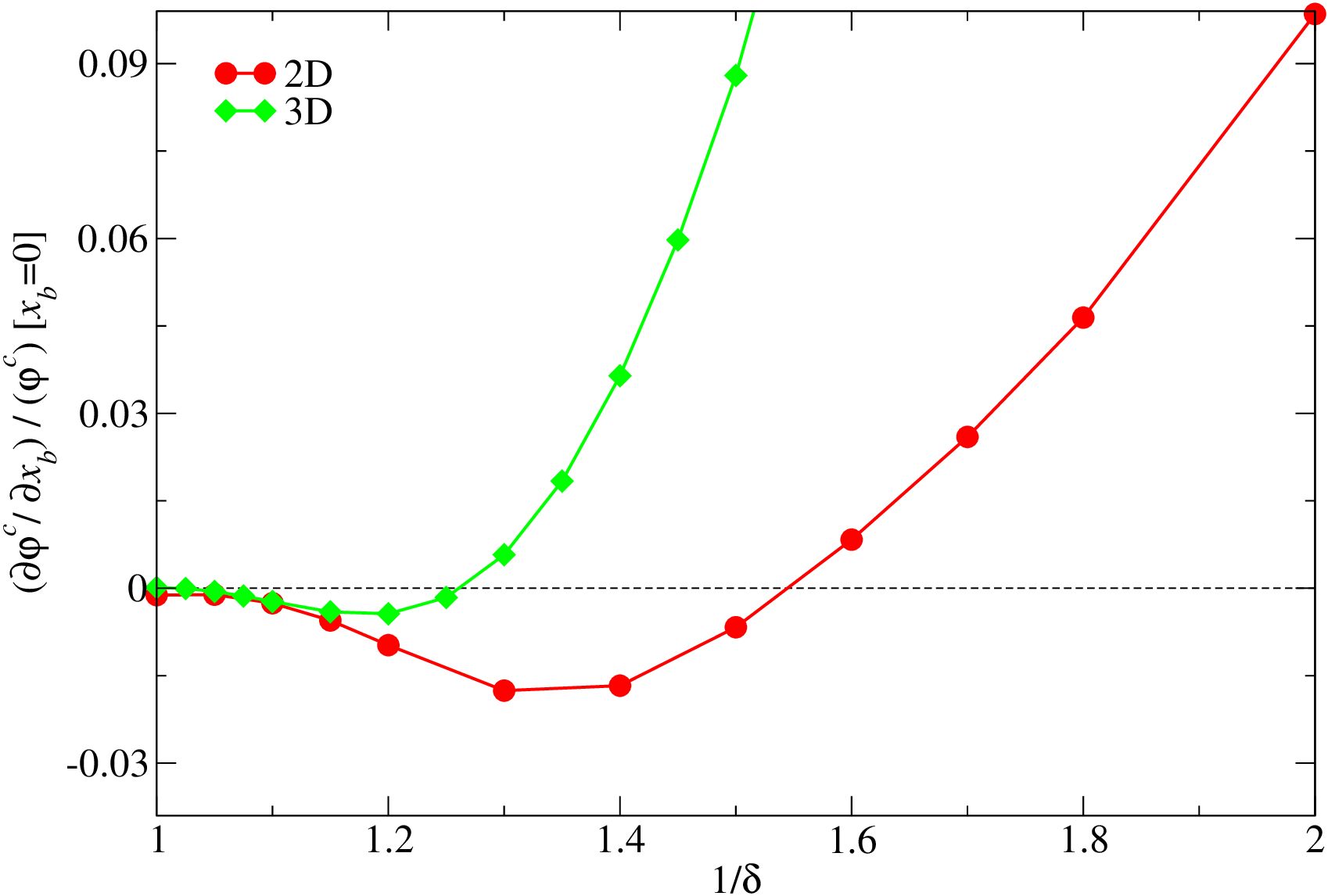}
\caption{(Color online) Slopes of the GTLs at $x_b=0$.}
\label{Fig.2}
\end{figure}

Figure~\ref{Fig.1} shows normalized slopes of the GTLs at $x_s=0$ as
functions of $1/\delta$ for the binary 2D and 3D HSM models.
Because $\delta=1$ 
represents a one component system, the slopes have to be zero at this point.
While the numerical results for the 3D model clearly support this statement, the numerical data
for the 2D model at $\delta=1$ slightly deviate from zero (see also Fig.~~\ref{Fig.2}). This, however, is an artifact due to
the numerically calculated static structure factors in 2D. For the 3D model, we have used the analytical solution of
Eqs.~(\ref{oz}) and (\ref{py}) to calculate the static input for MCT which has led to
a better self-consistency at $\delta=1$ then for the 2D model.
For $\delta$ close to unity, the slopes become negative which means that the presence of a small
concentration of the smaller particles stabilizes the glass. After exhibiting a minimum at $\delta_-$, the slopes become
zero again at $\delta_0$ and remain positive for $\delta<\delta_0$. Here the presence of the smaller particles stabilizes the liquid
which is nothing but the well known plasticization effect. Upon further decreasing of $\delta$, the slopes exhibit a maximum at $\delta_+$
and indicate a monotonic decay for asymptotically small $\delta$. For the 2D model, this decay
is more stretched then for the 3D case.

For the 3D model, we observe a continuous transition of the tagged particle NEPs $(\boldsymbol{F}^{c,(1)})_k^{ss}$ (see Appendix \ref{tagged})
to zero by approaching $\delta\approx1/4$ from above. This indicates a delocalization transition of the smaller
spheres in the glass formed by the bigger ones \cite{Bosse87,Bosse91,Bosse95}.
Such a transition is strongly influenced by
a $1/k^2$-divergence of the memory kernel for the tagged-particle correlators at $k=0$
\cite{Leutheusser83}. This singularity reflects the fact that, inside a fluid, the momentum of a single
tagged-particle is not conserved.
Although the evaluation of Eq.~(\ref{slope}) at $x_s=0$ requires $(\boldsymbol{F}^{c,(1)})_k^{ss}$ as input,
the qualitative $x_s$-dependence of $\varphi^c$ should not be influenced by this problem. Nevertheless,
we show the corresponding data for $1/\delta>3.5$ in Fig.~\ref{Fig.1} with open symbols.
However, these data show the same qualitative behavior as the corresponding ones for the 2D model.
For our choice of the lower cutoff for $k$, the MCT model does not yield a delocalization transition in 2D,
even if we use the PY result for $R_s=0$ as static input.
This, however, is an artifact due to the singularity of the tagged-particle memory kernel at $k=0$. Again, the qualitative $x_s$-dependence of $\varphi^c$ should
not be influenced.

Figure~\ref{Fig.2} shows normalized slopes of the GTLs at $x_b=0$.
For $\delta$ close to unity, the presence of a small concentration of the bigger
particles leads to a stabilization of the glass.
The slope vanishes at $\delta'_0>\delta_0$.
A strongly increasing plasticization effect occurs for smaller $\delta$.

\begin{figure}
\includegraphics[width=1\columnwidth]{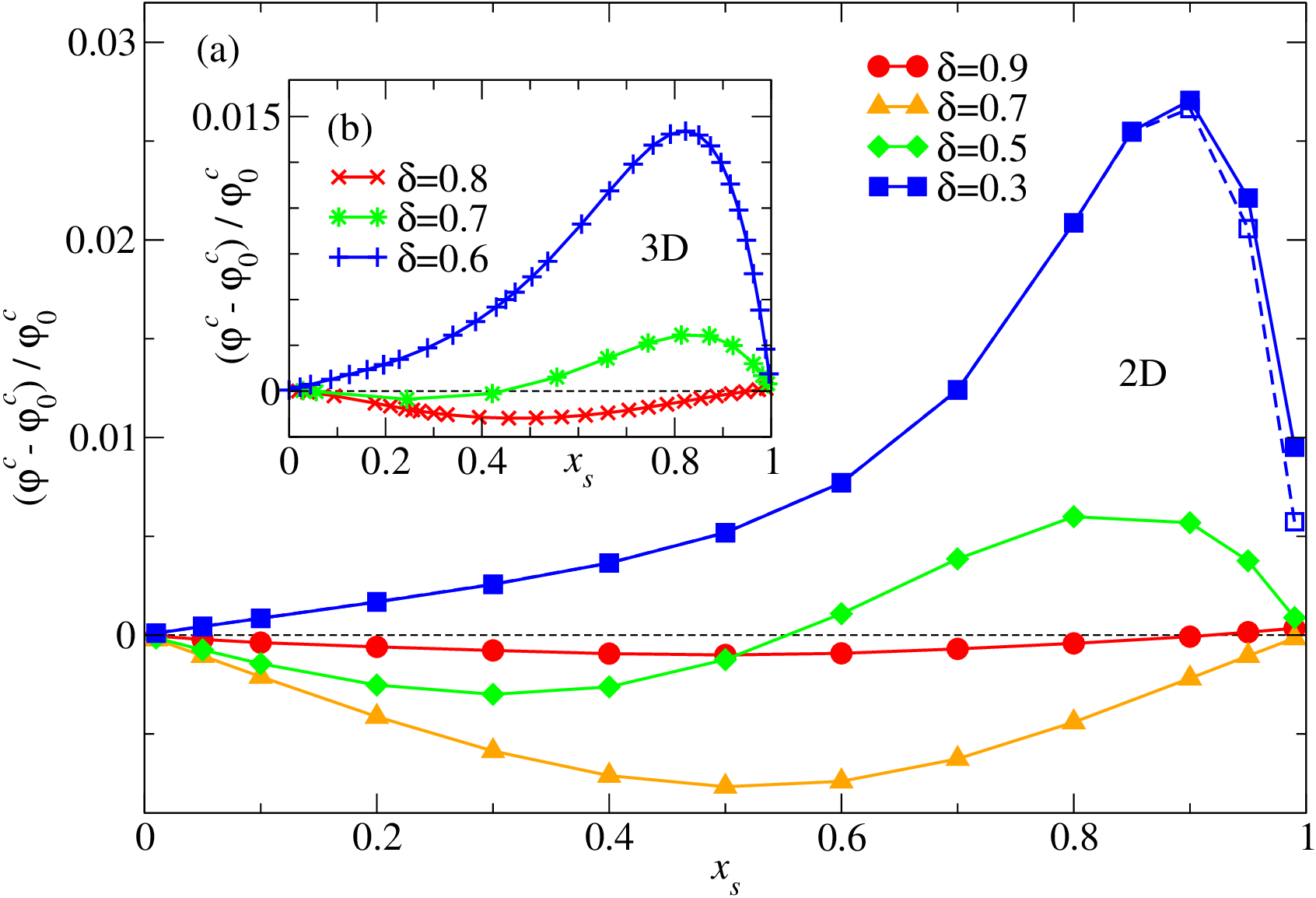}
\caption{(Color online) (a)~Relative variation of the glass transition lines for the binary HSM model in 2D.
The open squares calculated with $K=400$ grid points (instead of $K=250$) give an estimate for the error due to the high wave number cutoff.
(b)~Relative variation of the glass transition lines for the binary HSM model in 3D using the numerical data of G\"otze and
Voigtmann \cite{Voigtmann03} calculated with $\Delta k=0.4$ and $K=200$.}
\label{Fig.3}
\end{figure}

Apart from the problems discussed above,
the results shown in Figs.~\ref{Fig.1} and \ref{Fig.2} allow to predict the shape of the GTLs.
Both $x_s=0$ and $x_b=0$ define one component models with the same critical packing
fraction $\varphi_0^c$. Hence the GTLs show a single minimum for $\delta'_0<\delta<1$,
exhibit a minimum followed by a maximum (S-shape) for intermediate $\delta_0<\delta<\delta'_0$,
and show a single maximum for smaller $\delta<\delta_0$. Here we assumed that two or more minima (maxima) do not occur.
Figure~\ref{Fig.3}~(a) shows the relative variation $(\varphi^c-\varphi_0^c)/\varphi_0^c$ of the GTLs for the binary 2D HSM model. Results of G\"otze and Voigtmann for the 3D model \cite{Voigtmann03} are shown in Fig.~\ref{Fig.3}~(b).
The $\delta$-dependence of these GTLs agrees with the $\delta$-dependence predicted from the slopes at
$x_s=0$ and $x_b=0$. Particularly, the S-shapes of the GTLs for $\delta_0<\delta<\delta'_0$ are reproduced.

All our results predict the following trend: Compared to the 3D model, the stabilization of the glass is much more
pronounced in the 2D model where the less pronounced plasticization effect only sets in at larger size ratios.
The maximum relative decrease of $\varphi^c$ occurring
at $\delta\cong0.7$ in 2D (see Fig.~\ref{Fig.3}~(a))
is about five times larger then the maximum
downshift of $\varphi^c$ in 3D which occurs at $\delta\cong0.8$
(see Fig.~\ref{Fig.3}~(b) and Fig.~2 in Ref.~\cite{Voigtmann03}).
Qualitatively, the binary hard disc liquid exhibits the same two mixing effects discussed in the introduction for the
hard sphere liquid.

A finer resolution of the slope $s^c(x_s,\delta)\equiv(\partial\varphi^c/\partial x_s)(x_s,\delta)$ in Fig.~\ref{Fig.1} for delta close to unity (see inset)
shows that
$(\partial s^c/\partial \delta)(x_s=0,\delta=1)=0$. The numerical data for the 3D model show this behavior more
clearly then the corresponding ones for the 2D model, for technical reasons mentioned above.
The resolution of Fig.~\ref{Fig.2} already exhibits that
$(\partial s^c/\partial \delta)(x_s=1,\delta=1)=0$. Therefore, $\varphi^c(x_s,\delta)$ at $x_s=0$ and $x_s=1$ is quadratic
in $\varepsilon=(1-\delta)$ for $\delta$ close to unity. Since Eq.~(\ref{symmetry}) has led to the same $\varepsilon$-dependence
at $x_s=1/2$, we conjecture that
\begin{equation}
\label{square}
\varphi^c(x_s,\delta)\sim(1-\delta)^2
\end{equation}
for \textit{all} $x_s$ and small size disparity. A numerical check for, e.g. $x_s=1/4$, has confirmed the validity of
Eq.~(\ref{square}) for $d=2$ and $d=3$.
Equations~(\ref{symmetry}) and (\ref{square}) imply
\begin{equation}
\label{symmetry2}
\varphi^c(x_s,\delta)\cong\varphi^c(1-x_s,\delta).
\end{equation}
Consequently, the GTLs become symmetric in $x_s$ with respect to $x_s=1/2$ in the limit of small size disparity. Then the \textit{maximum} enhancement of glass formation occurs at equimolar concentration $x_s=1/2$, excluding again the occurrence of
more then one minimum.

\begin{figure}
\includegraphics[width=1\columnwidth]{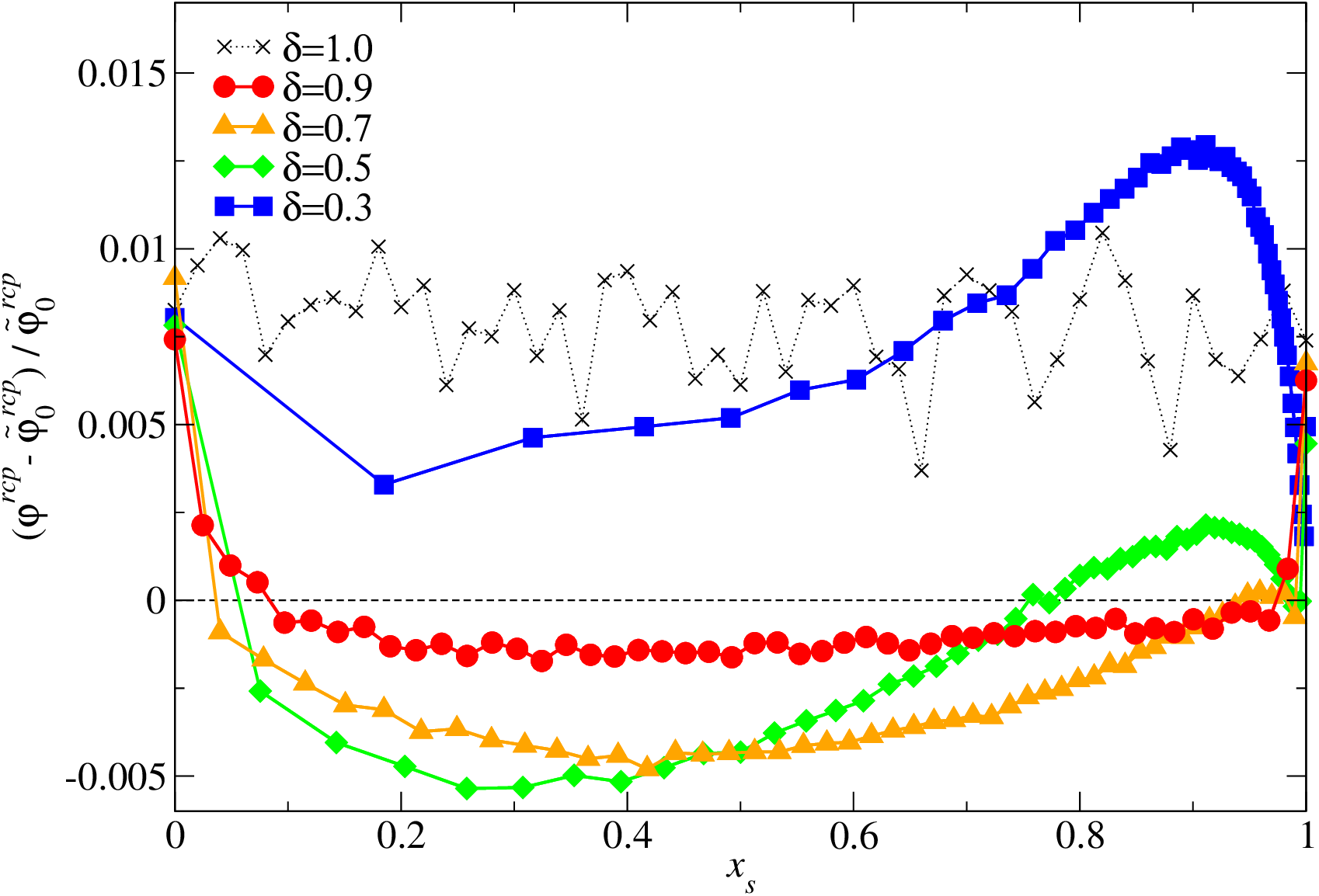}
\caption{(Color online) Relative variation of the random close packing fraction using the numerical data of
Okubo and Odagaki \cite{odagaki}. $\tilde{\varphi}_0^{rcp}$ is determined such that the relative variation vanishes
below but close to $x_s=1$.}
\label{Fig.4}
\end{figure}

Okubo and Odagaki \cite{odagaki} have numerically calculated
random close packing values \footnote{Since random close packing is not uniquely defined and depends on the procedure how it is realized, the comparison between $\varphi^c$ and $\varphi^{rcp}$ is on a qualitative level, only.} $\varphi^{rcp}$
of binary hard discs by use of a so-called
infinitesimal gravity protocol. Figure~\ref{Fig.4} presents their results for
$(\varphi^{rcp}-\tilde{\varphi}^{rcp}_0)/\tilde{\varphi}^{rcp}_0$.
$\tilde{\varphi}^{rcp}_0\cong0.8139$ is close but not identical to the averaged value $\varphi^{rcp}_0\cong0.82$ for monodisperse
hard discs. Despite the large numerical uncertainty at $x_s=0$, $x_s=1$ and $\delta=1$
(this might result from the fact that for monodisperse hard discs the applied procedure tends to build up
locally odered structures),
the data show a striking similarity to $(\varphi^c-\varphi_0^c)/\varphi_0^c$ (Fig.~\ref{Fig.3}~(a)).
The change from the single minimum shape to an S-shape and a maximum shape by decreasing $\delta$
is clearly reproduced by the random close packing result.

\subsection{Mixing scenarios}

\begin{figure}
\includegraphics[width=1\columnwidth]{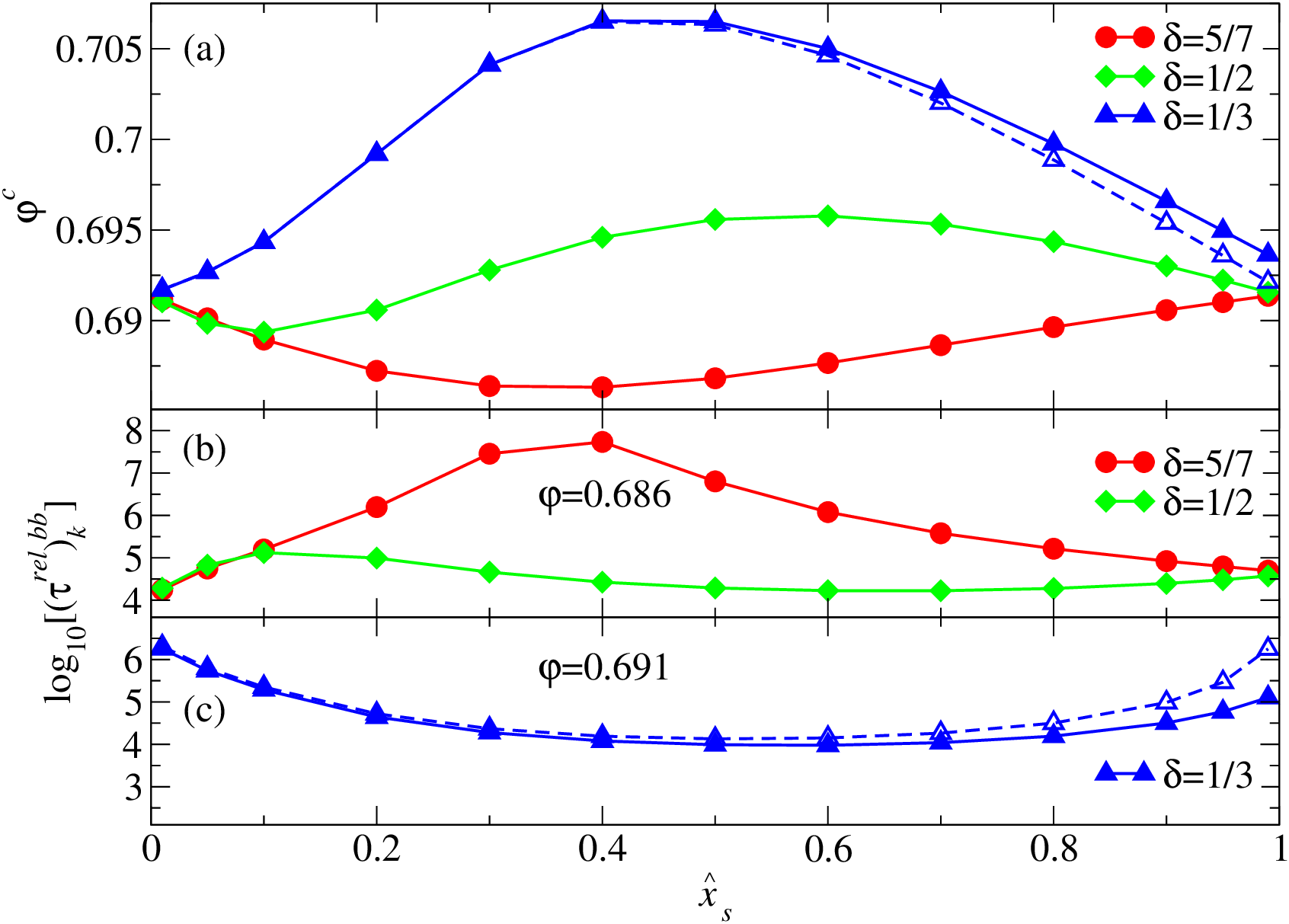}
\caption{(Color online)
(a) Glass transition lines for the binary HSM model in 2D, plotted as functions of the packing contribution
of the smaller particles $\hat{x}_s=\varphi_s/\varphi$.
(b)-(c) $\alpha$-relaxation times defined by
$\Phi_k^{bb}((\boldsymbol{\tau}^{rel})_k^{bb})=0.1(\boldsymbol{F}^c)_{k}^{bb}$
for the correlators of the big particles at $k=5.1909$ for fixed $\delta$ and $\varphi$ close below the corresponding GTLs
for the binary HSM model in 2D.
In both (a) and (c), the open triangles calculated with $K=400$ grid points (instead of $K=250$)  give an estimate for the error due to the high wave number cutoff.}
\label{Fig.5}
\end{figure}

In this section we will demonstrate that the mixing scenarios presented in Ref.~\cite{Voigtmann03} for binary hard spheres
are also observable for binary hard discs. For this purpose, we follow G\"otze and Voigtmann \cite{Voigtmann03} and
choose $\varphi$, $\delta$, and the packing contribution of the smaller particles
$\hat{x}_s=\varphi_s/\varphi$ as independent control parameters. In $d$ dimensions, we have
\begin{equation}
\label{xhat}
x_s=\frac{\hat{x}_s/\delta^d}{1+\hat{x}_s(1/\delta^d-1)}.
\end{equation}
As a direct analogon to Fig.~1 in Ref.~\cite{Voigtmann03}, Fig.~\ref{Fig.5}~(a) shows GTLs
for the binary HSM model in 2D, plotted as functions of $\hat{x}_s$ for
three representative values for $\delta$. The GTL for $\delta=5/7$ shows a single, clearly pronounced minimum,
the line for $\delta=1/2$ is S-shaped, and the GTL for $\delta=1/3$ exhibits a single maximum.

For both the hard sphere and the hard disc system the relative variation of $\varphi^c$ with concentration is of the order of one
percent or less (see Figs.~\ref{Fig.3} and \ref{Fig.5}~(a)). This can neither be observed by experiments nor by simulations.
As already stressed in Ref.~\cite{Voigtmann03},
the variation of $\varphi^c$ with, e.g. $\hat{x}_s$, may be reflected by a strong variation of the $\alpha$-relaxation time $(\boldsymbol{\tau}^{rel})_k^{\alpha\beta}$,
i.e. a variation of the characteristic time scale for the final decay of $\Phi_k^{\alpha\beta}(t)$ to zero in the liquid phase.
If $\varphi$ is fixed below but sufficiently close to $\varphi^c(\hat{x}_s,\delta)$,
i.e. if $\varphi$ is fixed such that there exists an interval in the $(\hat{x}_s,\delta)$-plane such that
$0<\varphi^c(\hat{x}_s,\delta)-\varphi\ll1$ is satisfied for all $(\hat{x}_s,\delta)$ within that interval, 
then the $\alpha$-relaxation time
$(\boldsymbol{\tau}^{rel})^{\alpha\beta}_k\sim(\varphi^c(\hat{x}_s,\delta)-\varphi)^{-\gamma(\hat{x}_s,\delta)}$ is extremely
sensitive to the variation of $(\hat{x}_s,\delta)$ within that interval.
Figure~\ref{Fig.5}~(b) shows $\alpha$-relaxation times defined by
$\Phi_k^{bb}((\boldsymbol{\tau}^{rel})_k^{bb})=0.1(\boldsymbol{F}^c)_{k}^{bb}$
for the unnormalized correlators of the big particles at $k=5.1909$ for fixed $\delta=5/7$, $1/2$ and fixed $\varphi=0.686$
below but close to
the corresponding GTLs for the binary HSM model in 2D for different packing contributions $\hat{x}_s$.
The qualitative $\hat{x}_s$-dependencies of the corresponding GTLs in Fig.~\ref{Fig.5}~(a)
are clearly reflected by the $\hat{x}_s$-dependencies of the $\alpha$-relaxation times.
$(\boldsymbol{\tau}^{rel})_k^{bb}$ shows a single maximum for $\delta=5/7$, and is S-shaped for $\delta=1/2$.
For $\delta=5/7$, $(\boldsymbol{\tau}^{rel})_k^{bb}$ varies by more then three decades.
Figure~\ref{Fig.5}~(c) shows $(\boldsymbol{\tau}^{rel})_k^{bb}$ at $k=5.1909$ for fixed $\delta=1/3$ and fixed $\varphi=0.691$
below but close to the corresponding GTL for the binary HSM model in 2D. The qualitative $\hat{x}_s$-dependence of the corresponding GTL in Fig.~\ref{Fig.5}~(a) is reflected by a single minimum in $(\boldsymbol{\tau}^{rel})_k^{bb}$.
Note that for this $\delta$ we had to choose a slightly larger value for $\varphi$ than for the two other examples shown in
Fig.~\ref{Fig.5}~(b) in order to clearly observe this effect. In contrast to this, Fig.~11 in Ref.~\cite{Voigtmann03}
exhibits all three scenarios for one common $\varphi$. In our 2D model, however, the minimum of $\varphi^c$ occurring for $\delta=5/7$ is more strongly pronounced then the corresponding one for $\delta=0.8$ in 3D shown in Fig.~1 in Ref.~\cite{Voigtmann03}.
This fact makes the choice of a common $\varphi$ for all three considered values of $\delta$ for the 2D model difficult.

\begin{figure}
\includegraphics[width=1\columnwidth]{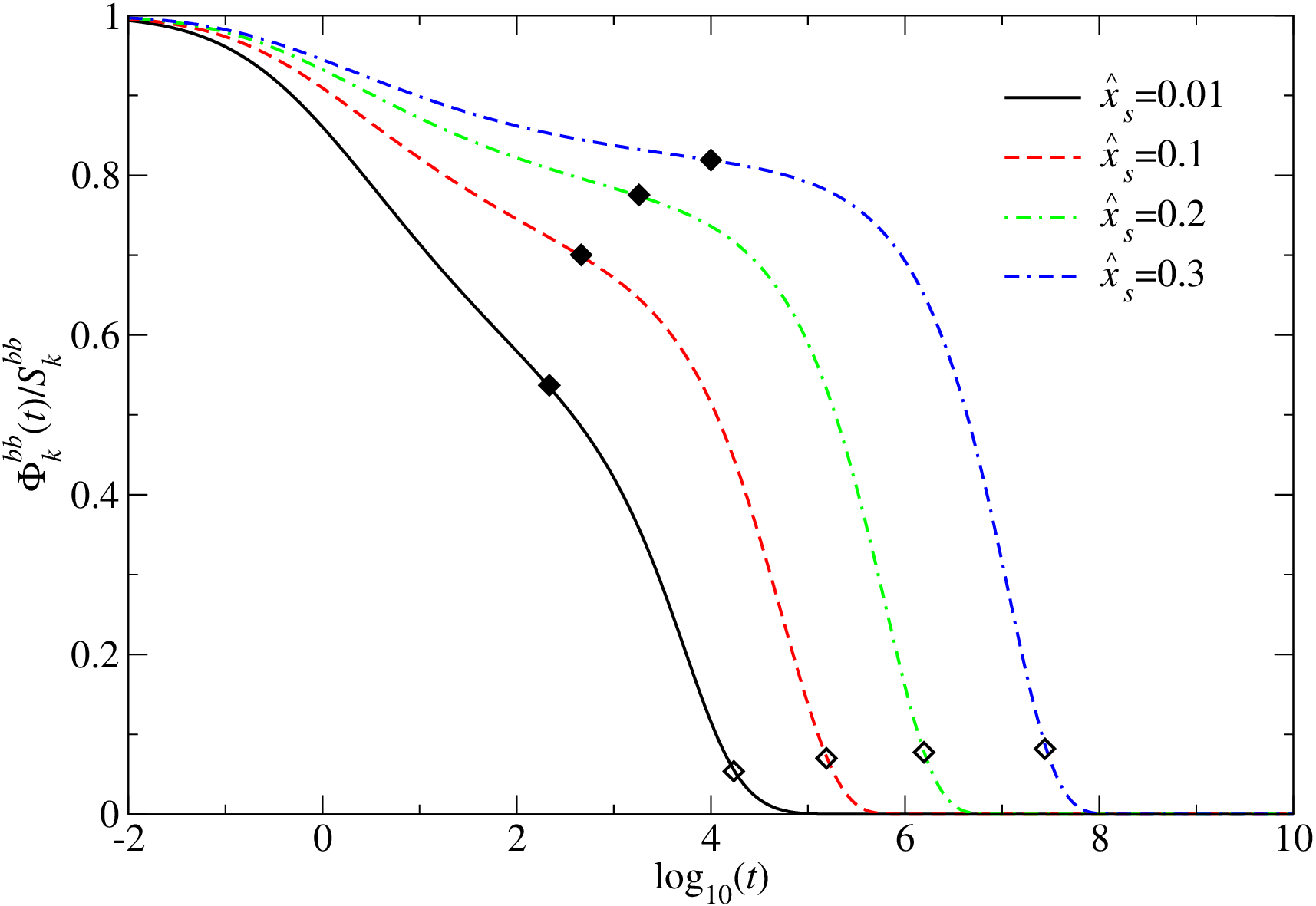}
\caption{(Color online) Normalized correlators of the big particles for the binary
HSM model in 2D at $\varphi=0.686$, $\delta=5/7$ and
$k=5.1909$ for different packing contributions $\hat{x}_s=\varphi_s/\varphi$. Filled diamonds mark the crossings of the normalized critical plateau values $(\boldsymbol{F}^c)^{bb}_{k}/(\boldsymbol{S}^c)^{bb}_{k}$. Open diamonds mark the crossings of the values
$0.1(\boldsymbol{F}^c)^{bb}_{k}/(\boldsymbol{S}^c)^{bb}_{k}$.}
\label{Fig.6}
\end{figure}

\begin{figure}
\includegraphics[width=1\columnwidth]{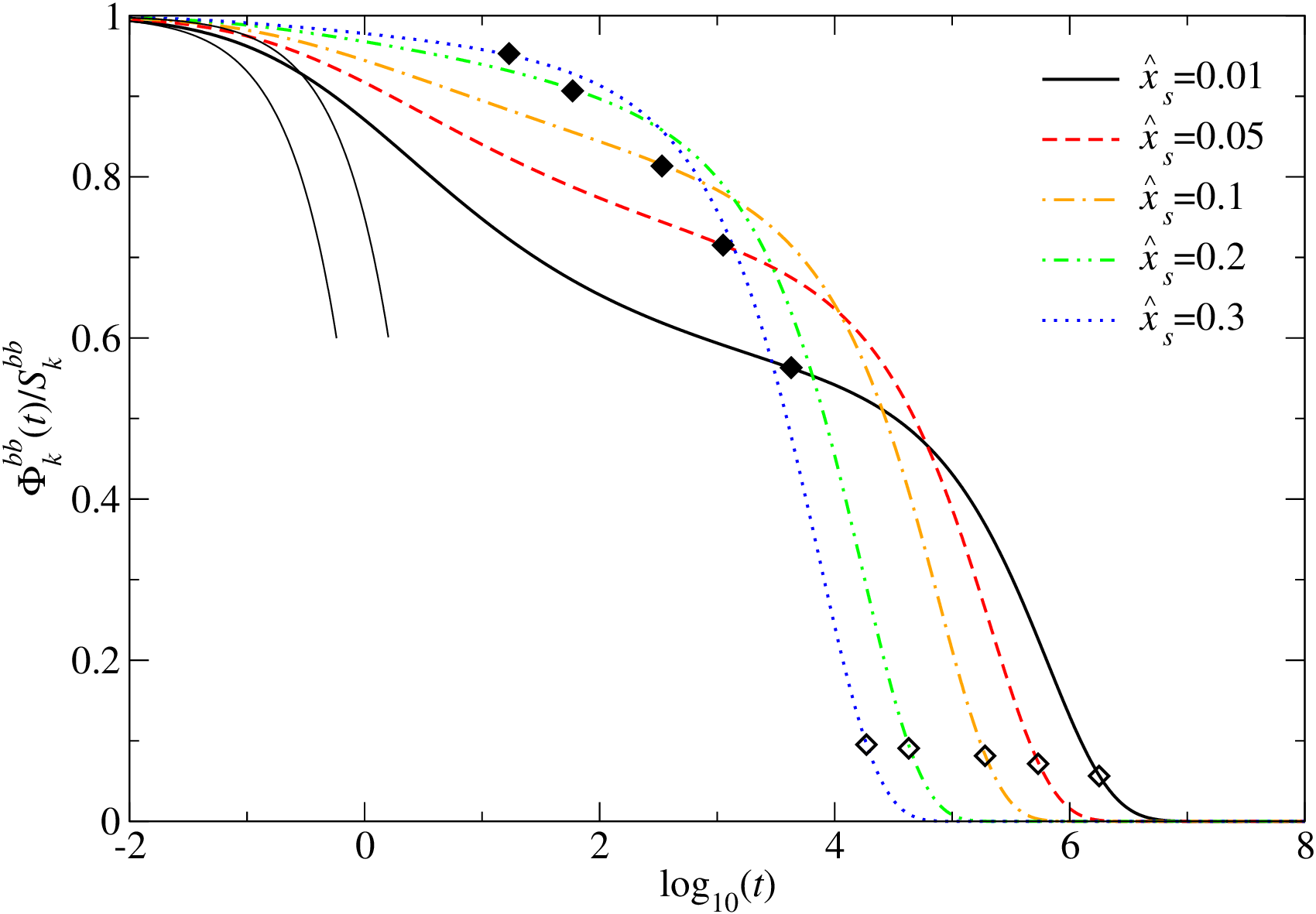}
\caption{(Color online) The same as Fig.~\ref{Fig.6} but for $\delta=1/3$ and $\varphi=0.691$. The thin lines show the short-time
asymptotes given by Eq.~(\ref{short-time}) for $\hat{x}_s=0.01$ and $\hat{x}_s=0.3$ (from left to right).}
\label{Fig.7}
\end{figure}

Let us also discuss some representative correlators in more detail.
As a direct analogon to the upper panel in Fig.~8 in Ref.~\cite{Voigtmann03},
Fig.~\ref{Fig.6} shows normalized correlators $\Phi_k^{bb}(t)/S_k^{bb}$ of the big particles for the binary
HSM model in 2D at fixed $\varphi=0.686$, $\delta=5/7$ and
$k=5.1909$ for different packing contributions $\hat{x}_s$ of the smaller discs.
Let $(\tilde{\boldsymbol{\tau}}^{rel})^{bb}_k$ be the characteristic time scale specified by $90\%$ of the decay from the normalized plateau
value $(\boldsymbol{F}^c)^{bb}_{k}/(\boldsymbol{S}^c)^{bb}_{k}$ to zero.
For the chosen value of $\delta$ the corresponding GTL shows a single minimum shape, see
Fig.~\ref{Fig.5}~(a). Hence, starting from the almost monodisperse system at $\hat{x}_s=0.01$ and
increasing the packing contribution of the smaller discs to $\hat{x}_s=0.3$ leads to a decrease of the distance $\varphi^c(\hat{x}_s,\delta)-\varphi$ from the GTL. This fact is reflected by an increase of $(\tilde{\boldsymbol{\tau}}^{rel})^{bb}_k$ by more then three decades (see the open diamonds in Fig.~\ref{Fig.6}).

An analogous scenario to the upper panel of Fig.~9 in Ref.~\cite{Voigtmann03} is presented in Fig.~\ref{Fig.7}. It shows
normalized correlators $\Phi_k^{bb}(t)/S_k^{bb}$ of the big particles for the binary
HSM model in 2D at fixed $\varphi=0.691$, $\delta=1/3$ and
$k=5.1909$ for different packing contributions of the smaller discs. For the $\delta$ chosen here the corresponding GTL shows a
single maximum shape (see Fig.~\ref{Fig.5}~(a)).
Hence, starting at $\hat{x}_s=0.01$ and increasing the packing contribution of the smaller discs to $\hat{x}_s=0.3$ leads
to an increase of the distance $\varphi^c(\hat{x}_s,\delta)-\varphi$ from the GTL. As a result, $(\tilde{\boldsymbol{\tau}}^{rel})^{bb}_k$ decreases by about two decades (see the open diamonds in Fig.~\ref{Fig.7}).

\begin{figure}
\includegraphics[width=1\columnwidth]{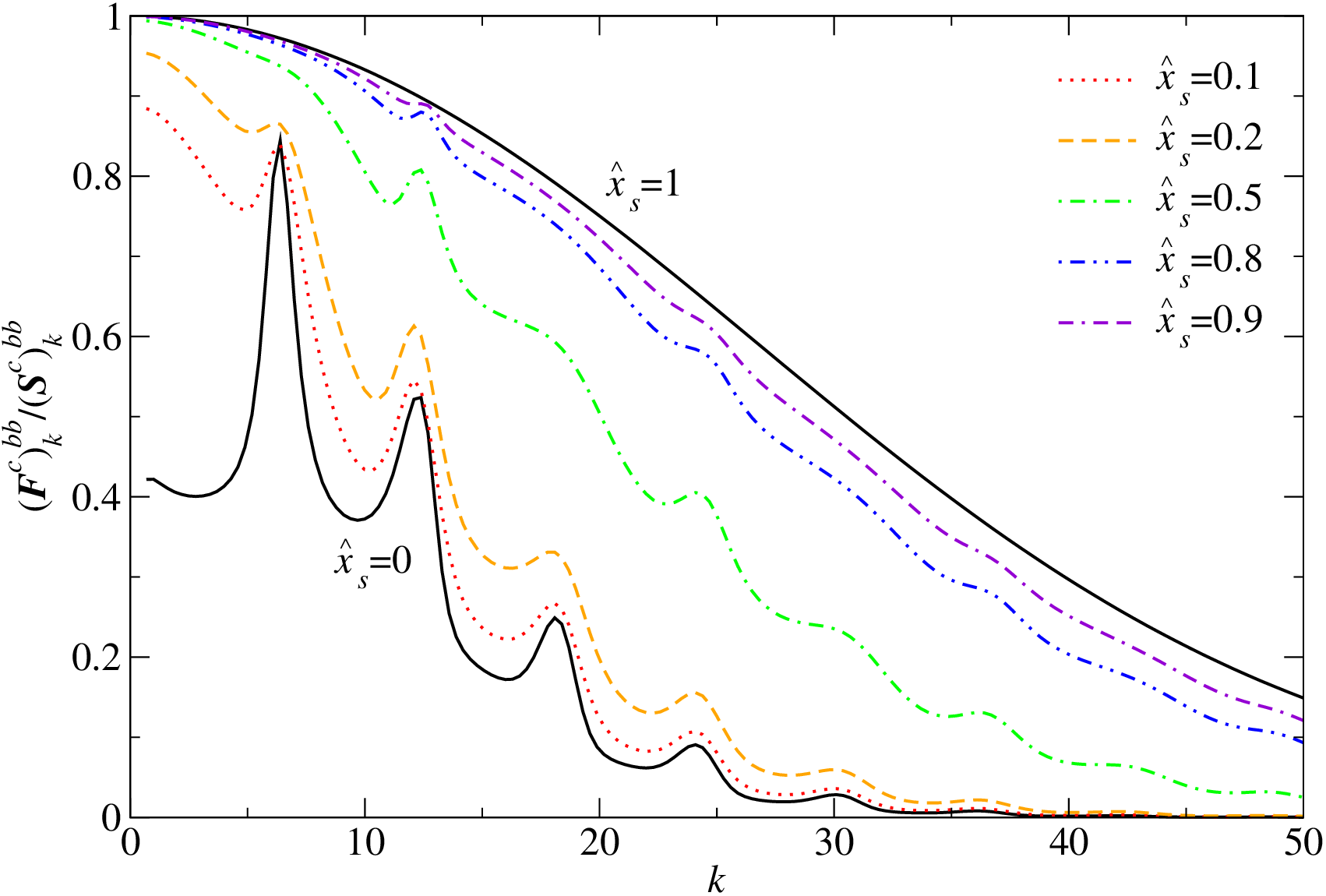}
\caption{(Color online) Normalized critical NEPs for the big particles at $\delta=1/2$ for the binary HSM model in 2D.}
\label{Fig.8}
\end{figure}

Two additional mixing effects (briefly mentioned in the introduction) were reported in Ref.~\cite{Voigtmann03} for the 3D model.
The first of these effects is the increase of the normalized critical NEPs (Debye-Waller factors)
$(\boldsymbol{F}^c)_k^{\alpha\alpha}/(\boldsymbol{S}^c)_k^{\alpha\alpha}$ upon increasing
$\hat{x}_s$ for almost all $k$ (related to an increase of the plateau values of the correlation functions for intermediate times).
The origin of this effect is explained in great detail in Ref.~\cite{Voigtmann03}. The 2D model shows similar behavior.
Here we restrict ourselves to a representative example. Figure~\ref{Fig.8} shows normalized critical NEPs
$(\boldsymbol{F}^c)_k^{bb}/(\boldsymbol{S}^c)_k^{bb}$ for the big particles at $\delta=1/2$ for the binary HSM model in 2D.
The data for $\hat{x}_s=0$ represent the Debye-Waller factors of a monodisperse system consisting of discs with diameter one,
while the result for $\hat{x}_s=1$ corresponds to the critical tagged particle NEPs (Lamb-M\"o{\ss}bauer factors) of a single
disc of diameter one inserted into a monodisperse system consisting of discs with diameter $\delta<1$. These Lamb-M\"o{\ss}bauer factors for all $k$ are larger then the corresponding Debye-Waller factors of the system of monodisperse discs with diameter one.
Provided that $(\boldsymbol{F}^c)_k^{bb}/(\boldsymbol{S}^c)_k^{bb}$ varies smoothly for all $0\leq\hat{x}_s\leq1$
(i.e. there are no multiple glassy states for the considered value of $\delta$), one obtains an increase of the Debye-Waller factors upon increasing $\hat{x}_s$ as an overall trend (see also the filled diamonds in Figs.~\ref{Fig.6} and \ref{Fig.7}).

The second remaining mixing effect is the slowing down of the initial part of the relaxation towards the plateau values for
the correlators of the big particles in the sense that $\Phi_k^{bb}(t)/S_k^{bb}$ versus $\log_{10}(t)$ becomes
flatter upon increasing $\hat{x}_s$. This effect is clearly visible is Figs.~\ref{Fig.6} and \ref{Fig.7}.
The authors of Ref.~\cite{Voigtmann03} conclude that the change of the short-time dynamics upon increasing $\hat{x}_s$ is
not sufficient to explain the observed effect. Figure~\ref{Fig.7} supports this statement. The shown short-time asymptotes
resulting from Eq.~(\ref{short-time}) for $\hat{x}_s=0.01$ and $\hat{x}_s=0.3$ fall already at $\log_{10}(t)\approx-1$
significantly below the corresponding correlators. Thus, the enormous flattening of the curves in the region $0<\log_{10}(t)<2$
can not be simply explained by the slowing down of the diffusion at short times.

Let us conclude at this point that we have found the same four mixing effects for binary hard discs as have been
reported for binary hard spheres in Ref.~\cite{Voigtmann03}. The subtle scenario in Fig.~\ref{Fig.7} is the result
of an interplay of three of these mixing effects. The increase of $\hat{x}_s$ leads first to both an increase of
the plateau values of the correlators at intermediate times and a slowing down of the initial part of the decay
toward these plateaus. However, the increase of $\hat{x}_s$ also leads to a decrease of the $\alpha$-relaxation times,
i.e. an enhancement of the final decay to zero,
and thus to a crossing of the correlators.

\section{Summary and conclusions}

In the present paper we have studied the influence of composition changes
on the glass transition for binary hard disc and hard sphere mixtures in the
framework of MCT.

By deriving Eq.~(\ref{slope1}), we have shown that the well-known separation parameter
not only describes the scaling of the NEPs in the glass
\cite{Goetze09, Franosch97}, but also describes the local variation of the GTLs to linear order.
For low concentration limits of one particle species we have evaluated the slopes of the GTLs, Eq.~(\ref{slope}),
by using a perturbation ansatz. With this we have introduced a new method which allows a fast prediction
of some qualitative properties of the GTLs.
Note that this method can be applied to
any MCT model with more then one control parameter. For instance, a similar analysis should be
possible for hard spheres with attractive potentials in the limit of vanishing attraction strength \cite{Dawson01},
or equivalently, for temperature going to infinity.
More generally, Eq.~(\ref{slope1}) holds for \textit{any} system which has at least
two control parameters and exhibits the generic $A_2$ bifurcation scenario \cite{Arnold}.

The direct comparison of the models in 2D and 3D show similar qualitative behavior.
Particularly, the same four mixing effects have been found as for hard spheres \cite{Voigtmann03}.
However, we have also found some differences.
The main difference is the fact that the extension of the glass regime due to mixing for size ratios
close to unity is
more strongly
pronounced in 2D then in 3D.

For small size disparity we have presented analytical and numerical evidence that the stabilization of the glassy state is
quadratic in $(1-\delta)$ and that the GTLs are almost symmetric with respect to their equimolar concentration $x_s=1/2$. At
this concentration the stabilization is maximal. These properties have not been noticed before.

Finally, we have shown that the qualitative $(x_s,\delta)$-dependence of $\varphi^c$ for some representative values of $\delta$
is identical to that of the random close packing $\varphi^{rcp}$. This is particularly true for the S-shape dependence for intermediate
values for $\delta$. The maximum shape variation of $\varphi^c$ which implies stabilization of the liquid state
and which has been related to entropic forces \cite{Voigtmann03,Fabbain99,Bergenholz99,Dawson01}
exists also for
$\varphi^{rcp}$ for smaller $\delta$. Since the random close packing procedure
of Ref.~\cite{odagaki} is a nonequilibrium process which maximizes the density locally,
it is not obvious that the stabilization effect is of entropic origin, at least for
$\delta$ not too small.

At this point we should also remember that $\varphi^{rcp}$ is not uniquely defined. For instance,
a subsequent shaking of the configurations produced by the infinitesimal gravity
protocol used in Ref.~\cite{odagaki} would typically lead to random structures at even higher densities.
Hence, one may ask whether the qualitative trends shown in Fig.~\ref{Fig.4} are reproducible by using
different procedures for calculating $\varphi^{rcp}$.
A different approach is the investigation of jamming transitions of hard discs or hard spheres.
Simulations on frictionless systems of repulsive spherical particles have
given evidence for a sharp discontinuity of the mean contact number $Z$ at a critical volume fraction $\varphi^{jam}$
\cite{Silbert02,Hern02,Hern03,Donev05}. These results are supported by experiments on binary photoelastic discs with
$\delta\approx0.86$ and $x_s=0.8$ \cite{Sperl07}. Recently, $\varphi^{jam}$ has been determined by St\"ark, Luding, and Sperl as function of $x_s$ for
different values for $\delta$, both by experiments on photoelastic discs and by corresponding computer
simulations \cite{Sperl09}. Their results clearly support all the qualitative features presented in Fig.~\ref{Fig.4},
whereby supporting the results shown in Fig.~\ref{Fig.3}.

Let us conclude with some open questions which are worth to be investigated in the future.
For the 3D model, higher order singularities (connected to the existence of multiple glassy sates)
occur below $\delta\approx0.4$ \cite{Voigtmann08}. The question, whether such transitions also exist
in 2D, requires a more detailed numerical study.
The consistency of our MCT results with the corresponding random close packing
data supports the quality of MCT in 2D. However, also a quantitative comparison of the
dynamical MCT results with molecular dynamics simulations is necessary.
A further step towards reality will be the study of MCT for binary discs including dipolar
interactions for which detailed experimental studies exist \cite{Koenig05}.

\acknowledgments

We thank M.~Bayer, T.~Franosch, M.~Fuchs, F.~H\"ofling, M.~Sperl and F.~Wey{\ss}er for stimulating discussions.
We especially thank W.~G\"otze for his valuable comments on this manuscript,
T.~Odagaki and T.~Okubo for providing the data for the random close packing of binary hard discs,
and Th.~Voigtmann for providing data for the 3D HSM model and for many helpful suggestions.

\appendix

\section{Evaluation of the slope in the weak mixing limit}

\label{weakmixing}

Here we will describe how to evaluate the slope of the GTL (Eq.~(\ref{slope})) at $x_s=0$. The procedure for $x_b=0$ is the same.
The corresponding formulae are obtained by interchanging the particle indices $b\leftrightarrow s$.
Let us further remark that the \textit{explicit} specialization on a certain model system occurs only on the
level of the static input for MCT. Thus, the  MCT formulae presented below can be directly translated and
applied to arbitrary binary mixtures such as soft sphere mixtures or binary discs including dipolar
interactions \cite{Koenig05}.

\subsection{Rewriting the mode coupling functional}

For the following, it is convenient to rewrite the mode coupling functional as
\begin{equation}
\label{fhat}
\boldsymbol{\mathcal{F}}=n\boldsymbol{x}^{-1}\hat{\boldsymbol{\mathcal{F}}}\boldsymbol{x}^{-1}
\end{equation}
where the elements of the matrix $\boldsymbol{x}$ are defined by $x_k^{\alpha\beta}=x_{\alpha}\delta_{\alpha\beta}$.
As can be read off from Eqs.~(\ref{memory})-(\ref{vertex}), $\hat{\boldsymbol{\mathcal{F}}}$
has a binlinear functional dependence on the matrix $\boldsymbol{c}$ of direct correlation functions,
and shows no further explicit dependence on the control parameters.
$\hat{\boldsymbol{\mathcal{F}}}$ can be considered as a special case of a more general functional $\tilde{\boldsymbol{\mathcal{F}}}$,
\begin{equation}
\hat{\boldsymbol{\mathcal{F}}}[\boldsymbol{X},\boldsymbol{Y}]=\tilde{\boldsymbol{\mathcal{F}}}[\boldsymbol{c},\boldsymbol{c};\boldsymbol{X},\boldsymbol{Y}],
\end{equation}
\begin{eqnarray}
\label{memorya}
\tilde{\mathcal{F}}_k^{\alpha\beta}[\boldsymbol{a},\boldsymbol{b};\boldsymbol{X},\boldsymbol{Y}]&=&\frac{\Omega_{d-1}}{(4\pi)^d}
\sum_{\alpha',\beta',\alpha'',\beta''}\int_0^\infty \mathrm{d} p\int_{|k-p|}^{k+p} \mathrm{d} q\nonumber\\
&&\times \tilde{V}^{\alpha\beta;\alpha'\beta',\alpha''\beta''}_{k;p,q}[\boldsymbol{a},\boldsymbol{b}]\nonumber\\
&&\times X_p^{\alpha'\beta'}Y_q^{\alpha''\beta''},
\label{bilineara}
\end{eqnarray}
\begin{equation}
\label{vertex_1a}
\tilde{V}^{\alpha\beta;\alpha'\beta',\alpha''\beta''}_{k;p,q}[\boldsymbol{a},\boldsymbol{b}]=
\frac{pq}{k^{d+2}}\tilde{v}_{kpq}^{\alpha\alpha'\alpha''}[\boldsymbol{a}]\tilde{v}_{kpq}^{\beta\beta'\beta''}[\boldsymbol{b}],
\end{equation}
\begin{equation}
\label{vertexa}
\tilde{v}^{\alpha\beta\gamma}_{kpq}[\boldsymbol{z}]=\frac{(k^2+p^2-q^2)z_p^{\alpha\beta}\delta_{\alpha\gamma}
+(k^2-p^2+q^2)z_q^{\alpha\gamma}\delta_{\alpha\beta}}{[
4k^2p^2-(k^2+p^2-q^2)^2]^{(3-d)/4}}.
\end{equation}
Hence, for \textit{fixed} $\boldsymbol{X}$ and $\boldsymbol{Y}$ 
and some arbitrary external control parameter $\xi_i$ we can write
\begin{equation}
\label{derivative}
(\partial\hat{\boldsymbol{\mathcal{F}}}/\partial{\xi_i})[\boldsymbol{X},\boldsymbol{Y}]=\hat{\boldsymbol{\mathcal{G}}}
[\boldsymbol{X},\boldsymbol{Y}],
\end{equation}
\begin{equation}
\hat{\boldsymbol{\mathcal{G}}}[\boldsymbol{X},\boldsymbol{Y}]=
\tilde{\boldsymbol{\mathcal{F}}}[\partial\boldsymbol{c}/\partial \xi_i,\boldsymbol{c};\boldsymbol{X},\boldsymbol{Y}]
+\tilde{\boldsymbol{\mathcal{F}}}[\boldsymbol{c},\partial\boldsymbol{c}/\partial \xi_i;\boldsymbol{X},\boldsymbol{Y}].
\end{equation}

\subsection{Derivatives of the separation parameter}

\subsubsection{General case}

For a general model system, the calculation of the slope of an arbitrary GTL (Eq.~(\ref{slope1})) requires the
calculation of a pair of derivatives of the separation parameter of the form $\partial\sigma/\partial(\Delta\xi_i)|_{\Delta\vec{\xi}=\vec{0}}$. Since $\sigma$ follows from $\tilde{\sigma}$ (Eq.~(\ref{sigma}))
by linearization around $\vec{\xi}^c$, we can write
\begin{equation}
\label{slope3}
\partial\sigma/\partial(\Delta\xi_i)|_{\Delta\vec{\xi}=\vec{0}}=\partial\tilde{\sigma}/\partial \xi_i|_{\vec{\xi}=\vec{\xi}^c}.
\end{equation}
Only, those quantities on the r.h.s of Eq.~(\ref{sigma}) without the superscript $c$ are differentiated. Then, all quantities in the resulting formula have to be taken at the critical point $\vec{\xi}^c$. For the following, we drop the superscript $c$ for
convenience. With Eqs.~(\ref{sigma}), (\ref{derivative}) and (\ref{slope3}) we obtain explicitly
\begin{eqnarray}
\label{general}
\partial\sigma/\partial(\Delta\xi_i)|_{\Delta\vec{\xi}=\vec{0}}&=&
n^{-1}(\partial n/\partial\xi_i)
(\hat{\boldsymbol{H}}|
\{\boldsymbol{S}
-\boldsymbol{F}\}\nonumber\\
&&\times n\boldsymbol{x}^{-1}\hat{\boldsymbol{\mathcal{F}}}[\boldsymbol{F},\boldsymbol{F}]\boldsymbol{x}^{-1}\{\boldsymbol{S}-\boldsymbol{F}\})\nonumber\\
&+&
(\hat{\boldsymbol{H}}|\{\boldsymbol{S}-\boldsymbol{F}\}\nonumber\\
&&\times n\boldsymbol{x}^{-1}\hat{\boldsymbol{\mathcal{G}}}[\boldsymbol{F},\boldsymbol{F}]\boldsymbol{x}^{-1}\{\boldsymbol{S}-\boldsymbol{F}\})\nonumber\\
&+&(\hat{\boldsymbol{H}}|\{\boldsymbol{S}-\boldsymbol{F}\}\boldsymbol{S}^{-1}\{\partial\boldsymbol{S}/\partial\xi_i\}
\nonumber\\
&&\times n\boldsymbol{x}^{-1}\hat{\boldsymbol{\mathcal{F}}}[\boldsymbol{F},\boldsymbol{F}]\boldsymbol{x}^{-1}\{\boldsymbol{S}-\boldsymbol{F}\})
\nonumber\\
&+&
(\hat{\boldsymbol{H}}|\{\boldsymbol{S}-\boldsymbol{F}\}\nonumber\\
&&\times n\boldsymbol{x}^{-1}\hat{\boldsymbol{\mathcal{F}}}[\boldsymbol{F},\boldsymbol{F}]\boldsymbol{x}^{-1}\{\partial\boldsymbol{S}/\partial\xi_i\})\nonumber\\
&+&
(\hat{\boldsymbol{H}}|\{\boldsymbol{S}-\boldsymbol{F}\}
n\{\partial\boldsymbol{x}^{-1}/\partial\xi_i\}\nonumber\\
&&\times
\hat{\boldsymbol{\mathcal{F}}}[\boldsymbol{F},\boldsymbol{F}]\boldsymbol{x}^{-1}\{\boldsymbol{S}-\boldsymbol{F}\})\nonumber\\
&+&
(\hat{\boldsymbol{H}}|\{\boldsymbol{S}-\boldsymbol{F}\}
n\boldsymbol{x}^{-1}
\hat{\boldsymbol{\mathcal{F}}}[\boldsymbol{F},\boldsymbol{F}]\nonumber\\
&&\times
\{\partial\boldsymbol{x}^{-1}/\partial\xi_i\}\{\boldsymbol{S}-\boldsymbol{F}\}).
\end{eqnarray}
Note that for a one-component model we have $\boldsymbol{x}=\boldsymbol{x}^{-1}=1$ and thus $\partial\boldsymbol{x}^{-1}/\partial\xi_i=0$. Let us further remark that the first scalar product on the r.h.s. of Eq.~(\ref{general}) is nothing but the well-known exponent parameter
$\lambda=(\hat{\boldsymbol{H}}|\{\boldsymbol{S}-\boldsymbol{F}\} \boldsymbol{\mathcal{F}} [\boldsymbol{F},\boldsymbol{F}]\{\boldsymbol{S}-\boldsymbol{F}\})$.

\subsubsection{Weak mixing limit}

We specialize Eq.~(\ref{general}) to evaluate Eq.~(\ref{slope}) at $x_s=0$.
Let us start with summarizing some important properties of $\boldsymbol{S}$, $\boldsymbol{F}$,
$\hat{\boldsymbol{\mathcal{F}}}$ and $\hat{\boldsymbol{\mathcal{G}}}$.
By definition, for $x_s\rightarrow0$ the elements $\boldsymbol{S}$ and $\boldsymbol{F}$ satisfy
\begin{equation}
\begin{array}{rcl}
\label{order1}
S_k^{\alpha\beta}=\mathcal{O}(x_s)&\text{if}&(\alpha,\beta)\neq(b,b),\\
F_k^{\alpha\beta}=\mathcal{O}(x_s)&\text{if}&(\alpha,\beta)\neq(b,b).
\end{array}
\end{equation}
Due to the Kronecker deltas is Eq.~(\ref{vertexa}), we also have
\begin{equation}
\label{order2}
\begin{array}{rcl}
\hat{\mathcal{F}}_k^{\alpha\beta}[\boldsymbol{F},\boldsymbol{F}]=\mathcal{O}(x_s)&\text{if}&(\alpha,\beta)\neq(b,b),\\
\hat{\mathcal{G}}_k^{\alpha\beta}[\boldsymbol{F},\boldsymbol{F}]=\mathcal{O}(x_s)&\text{if}&(\alpha,\beta)\neq(b,b).
\end{array}
\end{equation}
For the following,
we assume Taylor expansions for $n$, $\boldsymbol{c}$, $\boldsymbol{S}$, $\boldsymbol{F}$, $\boldsymbol{H}$ and $\hat{\boldsymbol{H}}$
in powers of $x_s$ around $x_s=0$ of the form
\begin{equation}
\label{taylor}
\boldsymbol{Z}=\boldsymbol{Z}^{(0)}+x_s\boldsymbol{Z}^{(1)}+\mathcal{O}(x_s^2).
\end{equation}
Equation~(\ref{order1}) implies
\begin{equation}
\begin{array}{rcl}
\label{order3}
(\boldsymbol{S}^{(0)})_k^{\alpha\beta}=0&\text{if}&(\alpha,\beta)\neq(b,b),\\
(\boldsymbol{F}^{(0)})_k^{\alpha\beta}=0&\text{if}&(\alpha,\beta)\neq(b,b).
\end{array}
\end{equation}
The Taylor expansions of $\hat{\boldsymbol{\mathcal{F}}}[\boldsymbol{F},\boldsymbol{F}]$ and $\hat{\boldsymbol{\mathcal{G}}}[\boldsymbol{F},\boldsymbol{F}]$ needed below read explicitly
\begin{eqnarray}
\label{taylorF}
\hat{\boldsymbol{\mathcal{F}}}[\boldsymbol{F},\boldsymbol{F}]
&=&\hat{\boldsymbol{\mathcal{F}}}^{(0)}[\boldsymbol{F}^{(0)},\boldsymbol{F}^{(0)}]\nonumber+x_s\hat{\boldsymbol{\mathcal{G}}}^{(0)}[\boldsymbol{F}^{(0)},\boldsymbol{F}^{(0)}]\\
&+&2x_s\hat{\boldsymbol{\mathcal{F}}}^{(0)}[\boldsymbol{F}^{(0)},\boldsymbol{F}^{(1)}]+\mathcal{O}(x_s^2),
\end{eqnarray}
\begin{equation}
\label{taylorG}
\hat{\boldsymbol{\mathcal{G}}}[\boldsymbol{F},\boldsymbol{F}]=\hat{\boldsymbol{\mathcal{G}}}^{(0)}[\boldsymbol{F}^{(0)},\boldsymbol{F}^{(0)}]+\mathcal{O}(x_s),
\end{equation}
where the leading order functionals are given by
\begin{equation}
\label{zero_F}
\hat{\boldsymbol{\mathcal{F}}}^{(0)}[\boldsymbol{X},\boldsymbol{Y}]=\tilde{\boldsymbol{\mathcal{F}}}[\boldsymbol{c}^{(0)},\boldsymbol{c}^{(0)};\boldsymbol{X},\boldsymbol{Y}],
\end{equation}
\begin{equation}
\label{zero_G}
\hat{\boldsymbol{\mathcal{G}}}^{(0)}[\boldsymbol{X},\boldsymbol{Y}]=
\tilde{\boldsymbol{\mathcal{F}}}[\boldsymbol{c}^{(1)},\boldsymbol{c}^{(0)};\boldsymbol{X},\boldsymbol{Y}]
+\tilde{\boldsymbol{\mathcal{F}}}[\boldsymbol{c}^{(0)},\boldsymbol{c}^{(1)};\boldsymbol{X},\boldsymbol{Y}].
\end{equation}
Equation~(\ref{order3}) and the Kronecker deltas is Eq.~(\ref{vertexa}) imply
\begin{equation}
\label{order4}
\begin{array}{rcl}
(\hat{\boldsymbol{\mathcal{F}}}^{(0)}[\boldsymbol{F}^{(0)},\boldsymbol{F}^{(0)}])_k^{\alpha\beta}=0&\text{if}&(\alpha,\beta)\neq(b,b),\\(\hat{\boldsymbol{\mathcal{G}}}^{(0)}[\boldsymbol{F}^{(0)},\boldsymbol{F}^{(0)}])_k^{\alpha\beta}=0&\text{if}&(\alpha,\beta)\neq(b,b).
\end{array}
\end{equation}
A further important implication is the fact that $(\hat{\boldsymbol{\mathcal{F}}}^{(0)}[\boldsymbol{F}^{(0)},\boldsymbol{F}^{(1)}])_k^{\alpha\beta}$ is not dependent on
$(\boldsymbol{F}^{(1)})_k^{bb}$ if $(\alpha,\beta)\neq(b,b)$.

Now we consider the numerator in Eq.~(\ref{slope}). It follows from Eq.~(\ref{general}) by choosing $\xi_i=x_s$.
Let us focus on the scalar product in the first term on the r.h.s. of Eq.~(\ref{general}). The factors
$\{\boldsymbol{S}-\boldsymbol{F}\}\boldsymbol{x}^{-1}$, $n$,
$\hat{\boldsymbol{\mathcal{F}}}[\boldsymbol{F},\boldsymbol{F}]$ and
$\boldsymbol{x}^{-1}\{\boldsymbol{S}-\boldsymbol{F}\}$ have all well defined limits for $x_s\rightarrow0$
which can be calculated independently. Hence, the limit of the second argument of the considered scalar product also exists.
Thus, the $x_s\rightarrow0$ limit of the first argument of the scalar product, namely that of $\hat{\boldsymbol{H}}$, can be performed independently with $\hat{\boldsymbol{H}}^{(0)}$ as result. Because of Eqs.~(\ref{taylorF}) and (\ref{order4}),
the final result for the $x_s\rightarrow0$ limit of the first term on the r.h.s. of Eq.~(\ref{general})
depends only on the matrix elements with indices $(\alpha,\beta)=(b,b)$.
We can write the result explicitly as
$
\label{exp1}
\{n^{(1)}/n^{(0)}\}\lambda^{(0)}
$
where
\begin{eqnarray}
\label{exp1a}
\lambda^{(0)} &  = & n^{(0)} \sum_k(\hat{\boldsymbol{H}}^{(0)})^{bb}_{k}\{(\boldsymbol{S}^{(0)})^{bb}_{k}-(\boldsymbol{F}^{(0)})^{bb}_{k}\}\\
& & \times  (\hat{\boldsymbol{\mathcal{F}}}^{(0)}[\boldsymbol{F}^{(0)},\boldsymbol{F}^{(0)}])_k^{bb} \{(\boldsymbol{S}^{(0)})^{bb}_{k}-(\boldsymbol{F}^{(0)})^{bb}_{k}\}\nonumber
\end{eqnarray}
is nothing but the well known exponent parameter of
the corresponding monodisperse MCT model \cite{Goetze09,Franosch97}.
The second term on the r.h.s. of Eq.~(\ref{general}) can be discussed similarly, here Eqs.~(\ref{taylorG}) and (\ref{order4})
lead to
\begin{eqnarray}
\mu^{(0)} &  = & n^{(0)} \sum_k(\hat{\boldsymbol{H}}^{(0)})^{bb}_{k}\{(\boldsymbol{S}^{(0)})^{bb}_{k}-(\boldsymbol{F}^{(0)})^{bb}_{k}\}\\
& & \times  (\hat{\boldsymbol{\mathcal{G}}}^{(0)}[\boldsymbol{F}^{(0)},\boldsymbol{F}^{(0)}])_k^{bb} \{(\boldsymbol{S}^{(0)})^{bb}_{k}-(\boldsymbol{F}^{(0)})^{bb}_{k}\}.\nonumber
\end{eqnarray}
The treatment of the remaining terms in Eq.~(\ref{general}) is somewhat more tedious.
For this purpose we write the matrix products occurring as second arguments of the scalar products
explicitly in components. By using Eqs.~(\ref{order1}) and (\ref{order2})
we realize that all the inverse powers of $x_s$ stemming from $\boldsymbol{x}^{-1}$ and its derivative with
respect to $x_s$ can be compensated by other factors which are of the order $x_s$.
Hence, the $x_s\rightarrow 0$ limits for all matrix products occurring as second arguments of the scalar products
exist. Thus, for each scalar product, the $x_s\rightarrow0$ limit of $\hat{\boldsymbol{H}}$ can be performed independently
yielding $\hat{\boldsymbol{H}}^{(0)}$. The final result for numerator in Eq.~(\ref{slope})
evaluated at $x_s=0$
can be written as
\begin{eqnarray}
\label{slope4}
\partial\tilde{\sigma}/\partial x_s|_{x_s=0 }& = & \{2+n^{(1)}/n^{(0)}\}\lambda^{(0)}
+ \mu^{(0)}\nonumber\\
 & + &  (\hat{\boldsymbol{H}}^{(0)}|\{\hat{\boldsymbol{A}}^{(0)}+ \hat{\boldsymbol{B}}^{(0)}\}),
\end{eqnarray}
\begin{eqnarray}
(\hat{\boldsymbol{A}}^{(0)})_k^{\alpha\beta}&=& n^{(0)} \{(\boldsymbol{S}^{(0)})^{\alpha b}_{k}-(\boldsymbol{F}^{(0)})^{\alpha b}_{k}\}\\
& & \times (\hat{\boldsymbol{\mathcal{F}}}^{(0)}[\boldsymbol{F}^{(0)},\boldsymbol{F}^{(0)}])_k^{bb}(\boldsymbol{S}^{(1)})^{b\beta}_{k}
 \nonumber\\
&+ &  2 n^{(0)}\{(\boldsymbol{S}^{(0)})^{\alpha b}_{k}-(\boldsymbol{F}^{(0)})^{\alpha b}_{k}\}\nonumber\\
& & \times (\hat{\boldsymbol{\mathcal{F}}}^{(0)}[\boldsymbol{F}^{(0)},\boldsymbol{F}^{(1)}])_k^{bs}(\boldsymbol{F}^{(1)})^{s\beta}_{k}
\nonumber\\
&- &  2 n^{(0)}\{(\boldsymbol{S}^{(1)})^{\alpha s}_{k}-(\boldsymbol{F}^{(1)})^{\alpha s}_{k}\}\nonumber\\
& & \times (\hat{\boldsymbol{\mathcal{F}}}^{(0)}[\boldsymbol{F}^{(0)},\boldsymbol{F}^{(1)}])_k^{sb}
\{(\boldsymbol{S}^{(0)})^{b\beta}_{k}-(\boldsymbol{F}^{(0)})^{b\beta}_{k}\}
\nonumber\\
&- &  4 n^{(0)}\{(\boldsymbol{S}^{(1)})^{\alpha s}_{k}-(\boldsymbol{F}^{(1)})^{\alpha s}_{k}\}\nonumber\\
& & \times (\hat{\boldsymbol{\mathcal{F}}}^{(0)}[\boldsymbol{F}^{(0)},\boldsymbol{F}^{(1)}])_k^{ss}
\{(\boldsymbol{S}^{(1)})^{s\beta}_{k}-(\boldsymbol{F}^{(1)})^{s\beta}_{k}\}
\nonumber\\
&+ &  2 n^{(0)}\{(\boldsymbol{S}^{(1)})^{\alpha s}_{k}-(\boldsymbol{F}^{(1)})^{\alpha s}_{k}\}\nonumber\\
& & \times (\hat{\boldsymbol{\mathcal{F}}}^{(0)}[\boldsymbol{F}^{(0)},\boldsymbol{F}^{(1)}])_k^{ss}(\boldsymbol{S}^{(1)})^{s\beta}_{k},
\nonumber
\end{eqnarray}
\begin{equation}
\hat{\boldsymbol{B}}^{(0)}=\hat{\boldsymbol{K}}^{(0)}\hat{\boldsymbol{L}}^{(0)},
\end{equation}
\begin{equation}
\begin{array}{rcl}
(\hat{\boldsymbol{K}}^{(0)})_k^{bb}&=&1-(\boldsymbol{F}^{(0)})^{bb}_{k}/(\boldsymbol{S}^{(0)})^{bb}_{k},\\
(\hat{\boldsymbol{K}}^{(0)})_k^{bs}&=&(\boldsymbol{F}^{(0)})^{bb}_{k}(\boldsymbol{S}^{(1)})^{bs}_{k}/(\boldsymbol{S}^{(0)})^{bb}_{k}
-(\boldsymbol{F}^{(1)})^{bs}_{k},\\
(\hat{\boldsymbol{K}}^{(0)})_k^{sb}&=&0,\\
(\hat{\boldsymbol{K}}^{(0)})_k^{ss}&=&1-(\boldsymbol{F}^{(1)})^{ss}_{k},
\end{array}
\end{equation}
\begin{eqnarray}
\label{exp2}
(\hat{\boldsymbol{L}}^{(0)})_k^{\alpha\beta}&=& n^{(0)} (\boldsymbol{S}^{(1)})^{\alpha b}_{k}
(\hat{\boldsymbol{\mathcal{F}}}^{(0)}[\boldsymbol{F}^{(0)},\boldsymbol{F}^{(0)}])_k^{bb}\nonumber\\
& & \times
\{(\boldsymbol{S}^{(0)})^{b\beta}_{k}-(\boldsymbol{F}^{(0)})^{b\beta}_{k}\}
\nonumber\\
&+& 2 n^{(0)} (\boldsymbol{S}^{(1)})^{\alpha s}_{k}
(\hat{\boldsymbol{\mathcal{F}}}^{(0)}[\boldsymbol{F}^{(0)},\boldsymbol{F}^{(1)}])_k^{sb}\nonumber\\
& & \times
\{(\boldsymbol{S}^{(0)})^{b\beta}_{k}-(\boldsymbol{F}^{(0)})^{b\beta}_{k}\}
\nonumber\\
&+& 2 n^{(0)} (\boldsymbol{S}^{(1)})^{\alpha s}_{k}
(\hat{\boldsymbol{\mathcal{F}}}^{(0)}[\boldsymbol{F}^{(0)},\boldsymbol{F}^{(1)}])_k^{ss}\nonumber\\
& & \times
\{(\boldsymbol{S}^{(1)})^{s\beta}_{k}-(\boldsymbol{F}^{(1)})^{s\beta}_{k}\},
\end{eqnarray}
Due to the statement below Eq.~(\ref{order4}), the final result, Eq.~(\ref{slope4}), does not depend on $(\boldsymbol{F}^{(1)})_k^{bb}$. The term $2\lambda^{(0)}$ results from the $bb$-elements to the last two
scalar products in Eq.~(\ref{general}). The matrix $\hat{\boldsymbol{B}}^{(0)}$ represents the contribution of the
third term in Eq.~(\ref{general}) where $\hat{\boldsymbol{K}}^{(0)}$ is nothing but the $x_s\rightarrow 0$ limit
of $\{\boldsymbol{S}-\boldsymbol{F}\}\boldsymbol{S}^{-1}$ while $\hat{\boldsymbol{L}}^{(0)}$ is the corresponding limit
for the expression
$
\{\partial\boldsymbol{S}/\partial x_s\} n\boldsymbol{x}^{-1}\hat{\boldsymbol{\mathcal{F}}}[\boldsymbol{F},\boldsymbol{F}]\boldsymbol{x}^{-1}\{\boldsymbol{S}-\boldsymbol{F}\}
$. All remaining quantities are summarized to the matrix $\hat{\boldsymbol{A}}^{(0)}$.

Let us now consider the denominator in in Eq.~(\ref{slope}) which follows from Eq.~(\ref{general}) by choosing $\xi_i=\varphi$.
Since Eqs.~(\ref{order1}) and (\ref{order2}) remain valid if one replaces the corresponding quantities by their derivatives
with respect to $\varphi$ and since $\partial\boldsymbol{x}^{-1}/\partial \varphi=0$, the final result depends only on
the $bb$-matrix elements. Thus, the denominator in Eq.~(\ref{slope}) taken at $x_s=0$ follows directly
from the separation parameter of the monodisperse system. It is a positive constant.

\subsection{Slope of a critical line}

The explicit results above allow us to define a procedure for the calculation of the slope of a GTL at $x_s=0$.
It consists of five steps.

\subsubsection{Calculation of the critical point}

The \textit{first} step is the determination of the critical packing fraction $\varphi^c_0$ and the corresponding NEPs $(\boldsymbol{F}^{c,(0)})_k^{bb}$ by using the corresponding one-component model of big particles.
In the following, all quantities have to be taken at $\varphi=\varphi^c_0$, the critical packing fraction of the one-component system.
The denominator in Eq.~(\ref{slope}) taken at $x_s=0$ also follows directly
from the separation parameter of the monodisperse system. It is a positive constant which we calculate
by numerical differentiation, for simplicity.

\subsubsection{Calculation of the static structure}

$\boldsymbol{S}^{(0)}$ and $\boldsymbol{S}^{(1)}$ entering into $\partial\tilde{\sigma}/\partial x_s|_{x_s=0 }$ trough
Eqs.~(\ref{exp1a})-(\ref{exp2}) can be easily determined from $\boldsymbol{c}^{(0)}$ and $\boldsymbol{c}^{(1)}$
by using Eq.~(\ref{oz1}).
The result reads
\begin{equation}
\begin{array}{rcl}
(\boldsymbol{S}^{(0)})_k^{bb}&=&1/\{1-n^{(0)}(\boldsymbol{c}^{(0)})_k^{bb}\},\\
(\boldsymbol{S}^{(0)})_k^{bs}&=&0,\\
(\boldsymbol{S}^{(0)})_k^{ss}&=&0,
\end{array}
\end{equation}
\begin{equation}
\begin{array}{rcl}
(\boldsymbol{S}^{(1)})_k^{ss}&=&1,\\
(\boldsymbol{S}^{(1)})_k^{bs}&=&n^{(0)}(\boldsymbol{S}^{(0)})_k^{bb}(\boldsymbol{c}^{(0)})_k^{bs},\\
(\boldsymbol{S}^{(1)})_k^{bb}&=&\{(\boldsymbol{S}^{(0)})_k^{bb}\}^2\\
&&\times\{n^{(0)}[(\boldsymbol{c}^{(1)})_k^{bb}+(\boldsymbol{c}^{(0)})_k^{ss}]\\
&&-(n^{(0)}-n^{(1)})(\boldsymbol{c}^{(0)})_k^{bb}\\
&&-(n^{(0)})^2[(\boldsymbol{c}^{(0)})_k^{bb}(\boldsymbol{c}^{(0)})_k^{ss}-(\boldsymbol{c}^{(0)})_k^{bs}(\boldsymbol{c}^{(0)})_k^{sb}]\}\\
&-&(\boldsymbol{S}^{(0)})_k^{bb}\{1+n^{(0)}(\boldsymbol{c}^{(0)})_k^{ss}\}.
\end{array}
\end{equation}
Hence, in the \textit{second} step we have to determine $\boldsymbol{c}^{(0)}$ and $\boldsymbol{c}^{(1)}$.
Substituting $n=n^{(0)}+x_sn^{(1)}+\dots$,
$\boldsymbol{c}=\boldsymbol{c}^{(0)}+x_s\boldsymbol{c}^{(1)}+\dots$,
$\boldsymbol{h}$ analogous and $\boldsymbol{x}=\boldsymbol{x}^{(0)}+x_s\boldsymbol{x}^{(1)}$
into Eqs.~(\ref{oz}) and (\ref{py}) leads to the equations for $\boldsymbol{c}^{(n)}$
and $\boldsymbol{h}^{(n)}$ which have to be solved recursively. For $n=0$ and $n=1$, they read
\begin{equation}
\label{ozzero}
\boldsymbol{h}^{(0)} = \boldsymbol{c}^{(0)}+n^{(0)}\boldsymbol{c}^{(0)}\boldsymbol{x}^{(0)}\boldsymbol{h}^{(0)}
\end{equation}
with the zeroth order PY closure
\begin{equation}
\begin{array}{rcrr}
(\boldsymbol{h}^{(0)})^{\alpha\beta}(r)&=&-1,& r<(R_{\alpha}+R_{\beta}),\\
(\boldsymbol{c}^{(0)})^{\alpha\beta}(r)&=&0,& r>(R_{\alpha}+R_{\beta}),
\end{array}
\label{pyzero}
\end{equation}
and
\begin{eqnarray}
\boldsymbol{h}^{(1)}&=&\boldsymbol{c}^{(1)}+n^{(1)}\boldsymbol{c}^{(0)}\boldsymbol{x}^{(0)}\boldsymbol{h}^{(0)}
+n^{(0)}\{\boldsymbol{c}^{(1)}\boldsymbol{x}^{(0)}\boldsymbol{h}^{(0)}\nonumber\\
&&+\boldsymbol{c}^{(0)}\boldsymbol{x}^{(1)}\boldsymbol{h}^{(0)}+\boldsymbol{c}^{(0)}\boldsymbol{x}^{(0)}\boldsymbol{h}^{(1)}\}
\label{first_order_oz}
\end{eqnarray}
with the first order PY closure
\begin{equation}
\begin{array}{rcrr}
(\boldsymbol{h}^{(1)})^{\alpha\beta}(r)&=&0,& r<(R_{\alpha}+R_{\beta}),\\
(\boldsymbol{c}^{(1)})^{\alpha\beta}(r)&=&0,& r>(R_{\alpha}+R_{\beta}).
\label{py_first_order}
\end{array}
\end{equation}
Furthermore, we have $(\boldsymbol{x}^{(0)})_k^{bb}=1$, $(\boldsymbol{x}^{(1)})_k^{bb}=-1$, $(\boldsymbol{x}^{(1)})_k^{ss}=1$ and all other components are zero, and $n^{(0)}$ and $n^{(1)}$ are given by
\begin{equation}
\label{model_1}
\begin{array}{rcl}
n^{(0)}&=&(\varphi d)/(\Omega_d R_b^d),\\
n^{(1)}&=&n^{(0)}(1-(R_s/R_b)^d).
\end{array}
\end{equation}
Note that Eqs.~(\ref{pyzero}), (\ref{py_first_order}) and (\ref{model_1}) are the only \textit{explicitly} model dependent equations.
Hence, the procedure can be easily extended for both to arbitrary binary mixtures and to closure relations
different from PY. Let us further remark that $(\boldsymbol{c}^{(0)})_k^{bb}$ and $(\boldsymbol{h}^{(0)})_k^{bb}$
are nothing but the direct and total correlations functions for the one-component system of big particles.

\subsubsection{Calculation of the critical nonergodicity parameters}

\label{tagged}

Beside $(\boldsymbol{F}^{(0)})_k^{bb}$, the evaluation Eqs.~(\ref{exp1a})-(\ref{exp2})
requires also $(\boldsymbol{F}^{(1)})_k^{bs}$ and
$(\boldsymbol{F}^{(1)})_k^{ss}$ as input. It is straightforward to derive the equations for these quantities from the
fixed point equation $\boldsymbol{F}=\boldsymbol{\mathcal{I}}[\boldsymbol{F}]$ following from Eq.~(\ref{fixed_point}) by considering the limit $x_s\rightarrow 0$. We obtain
\begin{equation}
\label{nep2}
(\boldsymbol{F}^{(1)})_k^{ss}=1-\{1+2 n^{(0)}(\hat{\boldsymbol{\mathcal{F}}}^{(0)}[\boldsymbol{F}^{(0)},\boldsymbol{F}^{(1)}])_k^{ss} \}^{-1},
\end{equation}
\begin{eqnarray}
\label{nep3}
(\boldsymbol{F}^{(1)})_k^{bs}&=&
2 n^{(0)} (\boldsymbol{S}^{(1)})^{bs}_{k}(\hat{\boldsymbol{\mathcal{F}}}^{(0)}[\boldsymbol{F}^{(0)},\boldsymbol{F}^{(1)}])_k^{ss}\nonumber\\
&&\times\{1-(\boldsymbol{F}^{(1)})^{ss}_{k}\}
\nonumber\\
&+&
2 n^{(0)} (\boldsymbol{S}^{(0)})^{bb}_{k}(\hat{\boldsymbol{\mathcal{F}}}^{(0)}[\boldsymbol{F}^{(0)},\boldsymbol{F}^{(1)}])_k^{bs}\nonumber\\
&&\times\{1-(\boldsymbol{F}^{(1)})^{ss}_{k}\}
\nonumber\\
&+&
n^{(0)} (\boldsymbol{S}^{(0)})^{bb}_{k}(\hat{\boldsymbol{\mathcal{F}}}^{(0)}[\boldsymbol{F}^{(0)},\boldsymbol{F}^{(0)}])_k^{bb}\nonumber\\
&&\times\{(\boldsymbol{S}^{(1)})^{bs}_{k}-(\boldsymbol{F}^{(1)})^{bs}_{k}\}.
\end{eqnarray}
Since $(\boldsymbol{F}^{(0)})_k^{bb}$ have already been determined in the first step, Eq.~(\ref{nep2}) allows to calculate $(\boldsymbol{F}^{(1)})_k^{ss}$.
The r.h.s. of Eq.~(\ref{nep2}) does neither depend on $(\boldsymbol{F}^{(1)})_k^{bs}$ nor on $(\boldsymbol{F}^{(1)})_k^{bb}$.
The $(\boldsymbol{F}^{(1)})_k^{ss}$ are nothing but the tagged particle NEPs for a single small particle in the fluid of the big particles.
Finally, Eq.~(\ref{nep3}) allows us to calculate $(\boldsymbol{F}^{(1)})_k^{bs}$, since it is not dependent on $(\boldsymbol{F}^{(1)})_k^{bb}$ due to the statement below Eq.~(\ref{order4}).

\subsubsection{Calculation of the critical eigenvectors}

The evaluation Eqs.~(\ref{exp1a})-(\ref{exp2}) requires the zeroth order left eigenvector
$\hat{\boldsymbol{H}}^{(0)}$ as last input. For its unique determination, also the zeroth order
right eigenvector $\boldsymbol{H}^{(0)}$ is needed.
For $x_s\rightarrow 0$, Eq.~(\ref{stability}) reduces to
\begin{equation}
\boldsymbol{C}^{(0)}[\boldsymbol{Y}]=2n^{(0)}\hat{\boldsymbol{M}}^{(0)}\hat{\boldsymbol{\mathcal{F}}}^{(0)}[\boldsymbol{F}^{(0)},\boldsymbol{Y}]\hat{\boldsymbol{N}}^{(0)},
\end{equation}
\begin{equation}
\begin{array}{rcl}
(\hat{\boldsymbol{M}}^{(0)})_k^{bb}&=&(\boldsymbol{S}^{(0)})^{bb}_{k}-(\boldsymbol{F}^{(0)})^{bb}_{k},\\
(\hat{\boldsymbol{M}}^{(0)})_k^{bs}&=&(\boldsymbol{S}^{(1)})^{bs}_{k}-(\boldsymbol{F}^{(1)})^{bs}_{k},\\
(\hat{\boldsymbol{M}}^{(0)})_k^{sb}&=&0,\\
(\hat{\boldsymbol{M}}^{(0)})_k^{ss}&=&1-(\boldsymbol{F}^{(1)})^{ss}_{k},
\end{array}
\end{equation}
\begin{equation}
\begin{array}{rcl}
(\hat{\boldsymbol{N}}^{(0)})_k^{bb}&=&(\boldsymbol{S}^{(0)})^{bb}_{k}-(\boldsymbol{F}^{(0)})^{bb}_{k},\\
(\hat{\boldsymbol{N}}^{(0)})_k^{bs}&=&0,\\
(\hat{\boldsymbol{N}}^{(0)})_k^{sb}&=&(\boldsymbol{S}^{(1)})^{sb}_{k}-(\boldsymbol{F}^{(1)})^{sb}_{k},\\
(\hat{\boldsymbol{N}}^{(0)})_k^{ss}&=&1-(\boldsymbol{F}^{(1)})^{ss}_{k}.
\end{array}
\end{equation}
Now, $\boldsymbol{C}^{(0)}$ and the corresponding adjoint map $\hat{\boldsymbol{C}}^{(0)}$ allow us to calculate
the eigenvectors $\boldsymbol{H}^{(0)}$ and $\hat{\boldsymbol{H}}^{(0)}$ obeying the normalization,
\begin{equation}
\sum_k(\hat{\boldsymbol{H}}^{(0)})_k^{bb}(\boldsymbol{H}^{(0)})_k^{bb}=1,
\end{equation}
\begin{equation}
\sum_k(\hat{\boldsymbol{H}}^{(0)})_k^{bb}\{(\boldsymbol{H}^{(0)})_k^{bb}\}^2/
\{(\boldsymbol{S}^{(0)})_k^{bb}-(\boldsymbol{F}^{(0)})_k^{bb}\}=1.
\end{equation}
While for $\boldsymbol{H}^{(0)}$ only the $bb$-elements are nonvanishing, $\hat{\boldsymbol{H}}^{(0)}$
has nontrivial contributions for all particle indices.
$(\boldsymbol{H}^{(0)})_k^{bb}$ and $(\hat{\boldsymbol{H}}^{(0)})_k^{bb}$ are the eigenvectors for the one-component model
of big particles.

\subsubsection{Calculation of the slope}

Now, we have determined all quantities for the evaluation of Eqs.~(\ref{exp1a})-(\ref{exp2})
and are able to calculate the slope of the GTL by using Eqs.~(\ref{slope}) and (\ref{slope3}).

\end{document}